\documentclass[twocolumn,aps,amsmath,amssymb]{revtex4}
\usepackage{graphicx,bm,amsmath,amssymb,natbib}

\begin{document}

\title{Andreev spin qubits in multichannel Rashba nanowires}

\author{Sunghun Park}
\affiliation{Departamento de F\'{\i}sica Te\'orica de la Materia Condensada, Condensed Matter Physics Center (IFIMAC) and
Instituto Nicol\'as Cabrera, Universidad Aut\'onoma de Madrid, Spain}
\author{A. Levy Yeyati}
\affiliation{Departamento de F\'{\i}sica Te\'orica de la Materia Condensada, Condensed Matter Physics Center (IFIMAC) and
Instituto Nicol\'as Cabrera, Universidad Aut\'onoma de Madrid, Spain}

\date{\today}

\begin{abstract}
We theoretically analyze the Andreev bound states and their coupling to external radiation in superconductor-nanowire-superconductor Josephson junctions. We provide an effective Hamiltonian for the junction projected onto the Andreev level subspace and incorporating the effects of nanowire multichannel structure, Rashba spin-orbit coupling, and Zeeman field. 
Based on this effective model, we investigate the dependence of the Andreev levels and the matrix elements of the current operator on system parameters such as chemical potential, nanowire dimensions, and normal transmission. We show 
that the combined effect of the multichannel structure and the spin-orbit coupling gives rise to finite current matrix elements between odd states having different spin polarizations. Moreover, our analytical results allow to determine the appropriate
parameters range for the detection of transitions between even as well as odd states in circuit QED like experiments,
which may provide a way for the Andreev spin qubit manipulation.
\end{abstract}
%\pacs{}

\maketitle

\section{Introduction}
Hybrid semiconductor/superconductor (S) devices are becoming promising platforms to host topological superconductivity and thus
Majorana zero modes~\cite{Kitaev2001,Fu2008,Sau2010,Oreg2010,Lutchyn2010,Alicea2012}. The technological advances are allowing to perform fundamental studies of some more basic mesoscopic objects, the 
Andreev bound states (ABSs), which characterize any coherent weak link between two superconducting electrodes~\cite{Beenakker1991}. In this respect, the direct detection of the current carrying ABSs
through tunneling experiments~\cite{Pillet2010,Sueur2008} or through microwave spectroscopy~\cite{Zazunov2003,Kos2013,Bretheau2013,Bretheau2013_2,Janvier2015,Larsen2015,Lange2015} has constituted a great achievement whose extension to the
topological case is being pursued by several groups~\cite{Virtanen2013,Peng2016,Klees2017,Wiedenmann2016,Woerkom2016}. In particular, the microwave experiments of Ref.~\cite{Janvier2015} in metallic atomic contacts demonstrated the 
possibility of quantum manipulation of the ABSs, an approach which could be now extended to the hybrid semiconductor devices.

The experiments on atomic contacts have also demonstrated that {\it odd}-parity states, in which an excess quasiparticle is trapped within the subgap levels, 
are long lived and can get a significant population when the contact is close to perfect transmission and the phase difference approaches $\pi$~\cite{Zgirski2011}.
While this ``poisoning'' mechanism can become detrimental for all qubit proposals based on Majorana zero modes~\cite{Rainis2012} or ABSs~\cite{Olivares2014,Zazunov2014}, the spin
degree of freedom of long lived odd states can become itself the basis for another type of qubit. This is precisely the idea behind the Andreev spin qubit
(ASQ) proposal of Nazarov and coworkers~\cite{Chtchelkatchev2003,Padurariu2010}. 

The ASQ was first proposed to be realized in metallic atomic contacts with strong spin-orbit (like for instance using Pb) which would be responsible of the splitting 
of the spin states~\cite{Chtchelkatchev2003,Padurariu2010,Beri2008}. The hybrid nanowires now provide another possible platform for their realization due to their strong spin-orbit interaction and the
tunability of their conduction channels~\cite{Woerkom2016,Goffman2017}. While most experimental progress along this line has been achieved on high quality InSb/S and InAs/S hybrid nanowires~\cite{Mourik2012,Das2012,Albrecht2016,Zhang2016}, recent developments include also proximity coupled strips in two dimensional electron gases (2DEGs)~\cite{Suominen2017,Lee2017} which are promising
platforms in view of their potential scalability and tunability.
There are, however, a number of uncertainties which hinder the feasibility of this realization. In the first place, 
the single channel theory of ABSs in a Rashba nanowire predicts spin-degenerate states for zero Zeeman field and thus suggests that high fields are 
needed to remove this degeneracy~\cite{Cheng2012,Yokoyama2014}. On the other hand, this theory also predicts vanishing current matrix elements between the odd states thus making the visibility of their transitions in microwave experiments negligible.

The aim of the present work is to analyze theoretically the ABS structure and the current matrix elements relevant for the even and odd transitions 
in superconductor/nanowire/superconductor junctions. We show that even when only the lowest subband is occupied the influence of the higher 
subbands is essential both for the energy splitting of the ABSs at zero field~\cite{Reynoso2012,Murani2016} as well as for obtaining finite matrix elements between the odd states having different spin polarizations. Our approach allows us to obtain analytical results for all relevant quantities as a function of the model parameters such as length, 
width, and chemical potential in the nanowire region. Our analysis thus provide a powerful tool to guide the experiments in the development of ASQs
based on semiconducting nanowires.

The paper is organized as follows. 
In Sec.~\ref{Model}, we introduce a model describing multichannel nanowire Josephson junctions in the energy regime of single channel-transport, and obtain an effective Hamiltonian by projecting the full model onto the subspace spanned by subgap ABSs. By solving the Hamiltonian, we find analytical expressions for the Andreev energy levels. In Sec.~\ref{Current}, we define four distinct states corresponding to possible occupation-number configurations of ABSs. We refer to the state in which Andreev levels with negative energies are occupied as the ``ground state'', and the state being occupied by two quasiparticles with different spins as the ``excited state''. We term ``odd state'' for a single quasiparticle occupation of ABSs. We analytically derive the matrix elements of the current operator for even states (ground and excited states) and odd states. We further analyze their dependence on controllable parameters such as Zeeman field, chemical potential, and junction length. In Sec.~\ref{Exp}, we discuss the feasibility to observe the transitions between odd states in actual experiments.
Finally, in Sec.~\ref{Conclusion} we provide some concluding remarks.

\section{Model Hamiltonian}\label{Model}

\begin{figure}[!t]
\includegraphics[width=\columnwidth]{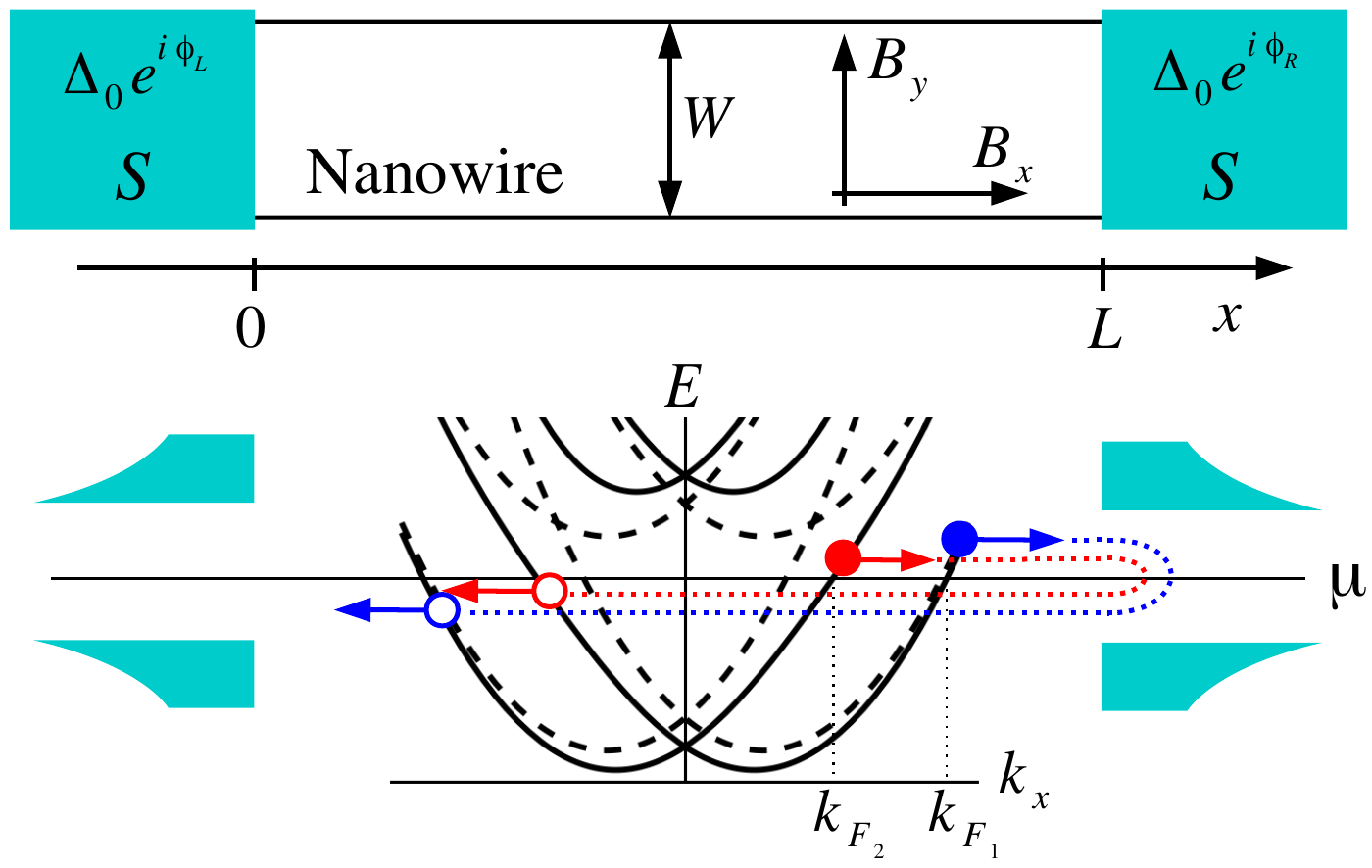}
\caption{Top: Schematic illustration of a quasi-one-dimensional nanowire proximity coupled to $s$-wave superconductors (S, cyan) 
forming a Josephson junction with a length $L$ and a width $W$. 
The nanowire supports many channels, Rashba-spin orbit coupling, a potential barrier, and magnetic fields $B_x$ and $B_y$. 
Bottom: Dispersion relation of the lowest two transverse subbands in the nanowire in the absence of magnetic fields, see Eq.~\eqref{Dispersion}. The case of $\eta=0$ is drawn by dashed lines and the $\eta \ne 0$ case by solid lines, where the subband coupling $\eta$ is given by Eq.~\eqref{Eta}. The second subbands with energy $E^{\perp}_{01}$ (Eq.~\eqref{T_energy}) which do not couple with the lowest ones through the spin-orbit coupling are not shown for clarity. 
Two co-propagating electrons (blue and red filled circles) with different Fermi velocities due to the finite $\eta$ are reflected as holes (blue and red empty circles), respectively, through Andreev reflection processes at $x=L$ (dotted lines). Multiple reflections at $x= 0$ and $L$ lead to the formation of Andreev levels. 
}
\label{Fig1:Setup}
\end{figure}

The model system we consider is schematically depicted in Fig.~\ref{Fig1:Setup}.
Electrons in a quasi-one-dimensional nanowire are confined in the $y$ and $z$ directions by an harmonic potential and are free to move in the $x$ direction. 
Two superconducting electrodes separated by a distance $L$ are proximity coupled to this nanowire forming a Josephson junction.
%connected through this nanowire form a Josephson junction and coherent 
%superpositions of the electrons and holes in the nanowire produced by Andreev 
%reflections at interfaces between the normal and superconducting regions give 
%rise to Andreev bound states. 
The Bogoliubov-de Gennes (BdG) Hamiltonian for this cylindrical Josephson junction is
%~\cite{Reynoso2012} 
\begin{align}
H_{\text{BdG}} = \left(H_0-\mu\right) \tau_z + H_R \tau_z + H_Z + H_S, \label{H_BdG}
\end{align}  
where $\mu$ is the chemical potential.
Here $H_0$ describes the quasi-one-dimensional nanowire given by
\begin{align}
H_0 = \frac{p^2_x+p^2_y+p^2_z}{2m} + U_{b}(x)+U_{c}(y,z), \label{H0}
\end{align}
where $m$ is the effective mass of the conduction electrons in the nanowire, $U_{b}(x) = U_0 \delta(x-x_0)$ represents the potential barrier at $x=x_0$ which allows to tune the junction transmission, and $U_c(y,z)= m \omega^2_0 (y^2+z^2)/2$ is the harmonic confinement potential where $\omega_0$ is the angular frequency. 
We define an effective diameter of the nanowire $W=2 \sqrt{\hbar/(m \omega_0)}$. 
We assume that a magnetic field is applied along the $x$ and $y$ directions, 
and that an electric field is present along the $z$ direction~\cite{Oreg2010,Zuo2017}.
The Rashba spin-orbit interaction $H_R$ and the Zeeman interaction $H_Z$ are given by 
\begin{align}
H_R &= -\alpha p_x\sigma_y + \alpha p_y \sigma_x, \label{SOI}\\
H_Z &= \frac{g\mu_B}{2}\left(B_x \sigma_x + B_y \sigma_y \right),
\end{align}
where $\alpha$ is the strength of the spin-orbit coupling and $B_x$ and $B_y$ are components of the applied magnetic field in the $x$ and $y$ directions, respectively. The Pauli matrices $\sigma_{x,y,z}$ and $\tau_{x,y,z}$ act in the spin and Nambu spaces, respectively. $H_S$ is the induced $s$-wave pairing potential due to the proximity effect,
\begin{align}
H_S = \Delta(x) \left(\text{cos}~\phi(x)\tau_x - \text{sin}~\phi(x)\tau_y \right) \label{Spairing}
\end{align}   
where the induced gap $\Delta(x)$ and the superconducting phase $\phi(x)$ are given by $\Delta(x) e^{i \phi(x)} = \Delta_0 e^{i \phi_L}$ at $x < 0$ and $ \Delta_0 e^{i \phi_R}$ at $x > L$. In the normal region of $0 < x < L$, $\Delta(x) = 0$.  
The superconducting phase difference is defined by $\phi = \phi_R - \phi_L$. Below, we assume that the potential barrier and the Zeeman field are weak so that we can treat $U_{b}(x)$ and $H_Z$ as perturbations.     

To make the discussion simpler, we define an effective one-dimensional (1D) BdG Hamiltonian by integrating out the $y$ and $z$ degrees of freedom. 
The sum of the kinetic and confinement terms in Eq.~\eqref{H0} associated with the $y$ and $z$ coordinates is $(p^2_y+p^2_z)/(2m)+U_c(y,z)$ which 
has the eigenvalues   
\begin{align}
E^{\perp}_{n_y n_z} = \hbar \omega_0 (n_y + n_z +1) = \frac{4 \hbar^2}{m W^2}(n_y + n_z +1),\label{T_energy}
\end{align}
where $n_y, n_z = 0, 1, 2,...$. The eigenstates $\phi^{\perp}_{n_y n_z s}(y,z)$ (with $s=\uparrow, \downarrow$) corresponding to the lowest two eigenvalues $\hbar \omega_0$ and $2 \hbar \omega_0$ are given by, 
\begin{align}
\phi^{\perp}_{00 s}(y,z) &= \frac{2}{\sqrt{\pi}W} e^{-2 (y^2 + z^2)/W^2} \chi_s, \nonumber\\
\phi^{\perp}_{10 s}(y,z) &= \frac{4 \sqrt{2} y}{\sqrt{\pi}W^2} e^{-2 (y^2 + z^2)/W^2} \chi_s, \nonumber\\
\phi^{\perp}_{01 s}(y,z) &= \frac{4 \sqrt{2} z}{\sqrt{\pi}W^2} e^{-2 (y^2 + z^2)/W^2} \chi_s,
\end{align}  
where $\chi_{\uparrow (\downarrow)} =  (1/\sqrt{2})(1, i (-i))^{T}$ are eigenstates of $\sigma_y$.
We note that the $\phi^{\perp}_{10 s}(y,z)$ and $\phi^{\perp}_{01 s}(y,z)$ are degenerate transverse modes with energy $2 \hbar \omega_0$.  
However, $\phi^{\perp}_{01 s}(y,z)$ do not couple to $\phi^{\perp}_{00 s'}(y,z)$ through the spin-orbit interaction 
\begin{align}
\int\int dy dz~ \phi^{\perp \dagger}_{00 s'}(y,z)~ H_R~ \phi^{\perp}_{01 s}(y,z) = 0,
\end{align} 
meaning that $\phi^{\perp}_{01 s}(y,z)$ do not contribute to the modification of the lowest subbands. 
By projecting $H_{\text{BdG}}$ onto the subspace spanned by the lowest two relevant transverse subbands, 
$\{\phi^{\perp}_{00\uparrow},\phi^{\perp}_{00\downarrow},\phi^{\perp}_{10\uparrow},\phi^{\perp}_{10\downarrow} \}$, followed by integrating out the $y$ and $z$ coordinates, we obtain 
\begin{gather}
H^{1D}_{\text{BdG}}~ \Psi(x) = \varepsilon~ \Psi(x), \nonumber\\
H^{1D}_{\text{BdG}} = \left(H'_0-\mu\right) \tau_z + H'_R \tau_z + H'_Z + H_S, \label{1DBdG}
\end{gather}
where $\Psi(x) = (\psi^{e}(x),\psi^{h}(x))^{T}$ with 
\begin{align}
\psi^{e}(x) &= (\psi^{e}_{0\uparrow}, \psi^{e}_{0\downarrow}, \psi^{e}_{1\uparrow}, \psi^{e}_{1\downarrow})^{T}, \nonumber\\
\psi^{h}(x) &= (\psi^{h}_{0\downarrow}, -\psi^{h}_{0\uparrow}, \psi^{h}_{1\downarrow}, -\psi^{h}_{1\uparrow})^{T},
\end{align}
where the subscripts $js$ on the $\psi^{e/h}_{js}$ denote the transverse quantum numbers $j=0,1$ and the spins $s=\uparrow, \downarrow$. 
$H'_0$, $H'_R$, and $H'_Z$ are the representations of $H_0$, $H_R$, and $H_Z$, respectively, in the subspace, 
\begin{align}
H'_0 &= \frac{p^2_x}{2m} + E^{\perp}_{+} + E^{\perp}_{-}\Sigma_z + U_{b}(x), \label{H'0}\\
H'_R &= -\alpha p_x \tilde{\sigma}_z + \eta \tilde{\sigma}_y \Sigma_y, \label{H'R} \\  
H'_Z &= \frac{g\mu_B}{2}\left(B_x \tilde{\sigma}_y + B_y \tilde{\sigma}_z \right),
\end{align}
where $E^{\perp}_{\pm} = (E^{\perp}_{00} \pm E^{\perp}_{10})/2$, the Pauli spin matrices $\tilde{\sigma}_{x,y,z}$ act in the spin space with basis $\{\chi_{\uparrow},\chi_{\downarrow}\}$, and $\Sigma_{x,y,z}$ are Pauli matrices acting on the space for the transverse degree of freedom.
The coefficient $\eta$ in Eq.~\eqref{H'R} describes the coupling between the different transverse subbands with opposite spins, and is given by 
\begin{align}
\eta &= \int dy dz~ \phi^{\perp \dagger}_{00 \downarrow}(y,z) 
\left(-i\hbar \alpha \frac{\partial}{\partial y} \sigma_x \right) \phi^{\perp}_{10 \uparrow}(y,z) \nonumber\\ 
&=\frac{\sqrt{2} \alpha \hbar}{W}. \label{Eta}
\end{align}

In the effective 1D model described by $H^{1D}_{\text{BdG}}$, the details of the system geometry such as 
dimensionality, subband states, and their energies 
enter through the parameters $E^{\perp}_{\pm}$ and $\eta$.    
If we construct a model Hamiltonian for a 1D nanowire starting from a 2DEG with a hard-wall confinement potential with width $W_{2d}$, the parameters are given by 
$E^{\perp}_{+}=5 \pi^2 \hbar^2/(4mW^2_{2d})$, $E^{\perp}_{-}=-3 \pi^2 \hbar^2/(4 mW^2_{2d})$, and 
$\eta = 8 \alpha \hbar/(3 W_{2d})$. As from the experimental point of view, quasi-one-dimensional wires can be made either from cylindrical nanowires or 2DEG heterostructures~\cite{Suominen2017}, we provide the results for Andreev levels and current matrix elements of Josephson junctions in a model for a 2DEG-based nanowire in App.~\ref{app:2Dnanowire}. We emphasize that although the specific forms of $E^{\perp}_{\pm}$ and $\eta$ depend on the dimensionality and confinement potential, the form of $H^{1D}_{\text{BdG}}$ in Eq.~\eqref{1DBdG} with $E^{\perp}_{\pm}$ and $\eta$ as parameters and the resulting analytical expressions, for instance, Eq.~\eqref{PBdG} below, are independent of such geometrical differences.        

We first examine $H'_0 + H'_R$ without the potential barrier $U_{b}(x)$.  In particular, we focus on the energy regime $E \lesssim E^{\perp}_{10}$ where spinful electrons move in a single channel (see Fig.~\ref{Fig1:Setup}). The dispersion relation in the energy regime is given by~\cite{Yokoyama2014,Reynoso2012,Murani2016} 
\begin{align}
E(k_x) = \frac{\hbar^2 k^2_x}{2m} + E^{\perp}_{+} 
- \sqrt{\left( E^{\perp}_{-} \mp \alpha \hbar k_x\right)^2 + \eta^2}, \label{Dispersion}
\end{align} 
and the Fermi velocities $v_{j=1,2}$ of the co-propagating electrons in the different spin subbands are 
\begin{align}
v_1 &= \frac{\hbar k^{e}_{x1}}{m} + \frac{\alpha \left(E^{\perp}_{-} - \alpha \hbar k^{e}_{x1} \right)}{\sqrt{\left( E^{\perp}_{-} - \alpha \hbar k^{e}_{x1}\right)^2 + \eta^2}},\nonumber\\
v_2 &= \frac{\hbar k^{e}_{x2}}{m}- \frac{\alpha \left(E^{\perp}_{-} + \alpha \hbar k^{e}_{x2} \right)}{\sqrt{\left( E^{\perp}_{-} + \alpha \hbar k^{e}_{x2}\right)^2 + \eta^2}},\label{Velocity}
\end{align}         
where $k^{e}_{xj}$ are wave vectors of the electrons.
If $\eta=0$, which means there is no mixing between the transverse subbands, 
we find that $v_1 = v_2$ because Eq.~\eqref{Velocity} reduces to $v_1=\hbar k^{e}_{x1}/m - \alpha$ and $v_2=\hbar k^{e}_{x2}/m + \alpha$ and Eq.~\eqref{Dispersion} gives $k^{e}_{x1} - k^{e}_{x2} = 2 m \alpha/\hbar$. If $\eta$ is finite, $v_1 \neq v_2$.
The eigenstates $\psi^{e}_{R,j=1,2} (\psi^{e}_{L,j=1,2})$ of electrons moving to the right (left) with the velocity $v_j$ are given by 
\begin{align}
\psi^{e}_{R,1} &= - \mathcal{T} \psi^{e}_{L,1} 
= \frac{e^{i k^{e}_{x1} x}}{\sqrt{|v_1|}} \left( \text{sin}\frac{\theta_1}{2},0,0, -\text{cos}\frac{\theta_1}{2}\right)^{T}, \nonumber\\
\psi^{e}_{R,2} &=  \mathcal{T} \psi^{e}_{L,2} 
= \frac{e^{i k^{e}_{x2} x}}{\sqrt{|v_2|}} \left( 0, \text{sin}\frac{\theta_2}{2},\text{cos}\frac{\theta_2}{2}, 0 \right)^{T}, \label{Eigenstates}
\end{align}
where $\mathcal{T} = -i \tilde{\sigma}_y \Sigma_0 \mathcal{C}$ is the time reversal operator where $\mathcal{C}$ indicates complex conjugation, and 
\begin{align}
\theta_1 &= \text{arccos}\left[\frac{1}{\alpha}\left(v_1 -\frac{\hbar k^{e}_{x1}}{m} \right) \right], \nonumber\\
\theta_2 &= \text{arccos}\left[\frac{1}{\alpha}\left(-v_2 +\frac{\hbar k^{e}_{x2}}{m} \right) \right].
\end{align}
For $\eta = 0$, $\theta_1 = \theta_2 = \pi$ and thus the spinors to the eigenstates have the forms $\psi^{e}_{R,1}, \psi^{e}_{L,2} \propto (1,0,0,0)^{T}$ and $\psi^{e}_{R,2}, \psi^{e}_{L,1} \propto (0,1,0,0)^{T}$, independent of the spin-orbit coupling and the momenta. The angles deviate from $\pi$ when $\eta$ is finite. In particular, in the limit $|E|, |E^{\perp}_{-}| \gg m \alpha^2, \eta$, they are expressed as 
\begin{align}
\text{cos}~\theta_{1(2)} \approx -1 + \frac{\eta^2}{2(E^{\perp}_{-} \mp \alpha \sqrt{2 m E})^2},\label{Approx_theta}
\end{align}
where the $-$ sign is for $\theta_{1}$ and $+$ for $\theta_{2}$.
We will see below that the the different Fermi velocities and different spin directions of two co-propagating electrons 
%, in combination with Zeeman effect, 
are a crucial ingredient for manipulating the Andreev levels.

In the following, we take into account the proximity-induced superconducting term given by Eq.~\eqref{Spairing}. 
%in the energy regime $\Delta_0 \ll \mu \lesssim E^{\perp}_{2}$ where a single 
%channel, whose dispersion relation is given by Eq.~\eqref{Dispersion}, 
%contribute to electron transport. 
%We consider a Josephson junction in the short limit that $\Delta_0 L / (\hbar v_j) \ll 1$. 
The corresponding BdG Hamiltonian is $\left(H'_{0} - \mu \right) \tau_z + H'_R\tau_z + H_S$. 
For further evaluation, we linearize the dispersion relation in Eq.~\eqref{Dispersion} in the normal region around the chemical potential $\mu$ far from the bottom of the subbands, 
\begin{align}
E^{e(h)}_{R,j} &= \mu \pm \hbar v_j \left( k^{e(h)}_{xj} -k_{F_j}\right), \nonumber\\
E^{e(h)}_{L,j} &= \mu \mp \hbar v_j \left( k^{e(h)}_{xj} + k_{F_j}\right), \label{LinearizedDispersion}
\end{align}
where the upper sign is for an electron and the lower for a hole. 
In the normal region without a potential barrier, coherent superpositions of electrons and holes produced by Andreev reflections at the interfaces between the normal and superconducting regions give rise to the ABSs. 
Perfect Andreev reflection at these interfaces connects time-reversed states. 
For instance, and electron with $E^{e}_{R,1}$ is converted to a hole with $E^{h}_{R,1}$, as illustrated in Fig.~\ref{Fig1:Setup}.
We also assume that the spinor parts of the eigenstates in Eq.~\eqref{Eigenstates} do not change significantly within the subgap energy regime $|\varepsilon| < \Delta_0$ so that $\theta_{j=1,2}$ are fixed as $\theta_{j}(k^{e}_{xj}) = \theta_{j}(k_{F_j})$. This is a good approximation provided that the subband separation is larger than the induced superconducting gap, $2 |E^{\perp}_{-}| \gg \Delta_0$.
By matching the wave functions at the interfaces, we obtain four normalized ABSs 
$\Psi_{j\lambda}(x)$ for $|\varepsilon| < \Delta_0$, where $j=1,2$ and $\lambda = \pm$.
The $\Psi_{1-}(x)$ and $\Psi_{2+}(x)$ have a component structure as 
\begin{align}
(\psi^{e}_{0\uparrow},0, 0, \psi^{e}_{1\downarrow},\psi^{h}_{0\downarrow}, 0, 0, -\psi^{h}_{1\uparrow})^{T},\label{Structure_1}
\end{align}
while $\Psi_{1+}(x)$ and $\Psi_{2-}(x)$ have 
\begin{align}
(0,\psi^{e}_{0\downarrow},\psi^{e}_{1\uparrow},0,0,-\psi^{h}_{0\uparrow},\psi^{h}_{1\downarrow},0)^{T},\label{Structure_2}
\end{align}
which are orthogonal to the states $\Psi_{1-}(x)$ and $\Psi_{2+}(x)$.  
Further details on the ABSs are given in App.~\ref{app:ABS}.  
The matching condition yields the following transcendental equation for the
Andreev level,
\begin{align}
\beta^{2} e^{i (k^{e}_{xj}-k^{h}_{xj})L + i \lambda \phi} = 1, \label{TranscendentalEqn}
\end{align}    
where $\beta = \varepsilon/\Delta_0 - i \sqrt{1-(\varepsilon/\Delta_0)^2}$. 
In the limit of either $\Delta_0 L / (\hbar v_j) \ll 1$ or $\varepsilon \ll \Delta_0$ and by using $e^{i (k^{e}_{xj}-k^{h}_{xj})L} = e^{i 2 \varepsilon L /(\hbar v_j)}$ from the linearized dispersion relation, the energy-phase relations, $\varepsilon_j(\phi)$ for $\Psi_{j+}(x)$ and $-\varepsilon_j(\phi)$ for $\Psi_{j-}(x)$, can be evaluated as
\begin{align}
\varepsilon_j(\phi) = \Delta_0~ \frac{\text{cos}(\phi/2)}{1+L_j~ \text{sin}(\phi/2)}, \label{EnergyPhaseRelation}
\end{align} 
where $L_j = \Delta_0 L/(\hbar v_j)$. The difference between $\varepsilon_1(\phi)$ and $\varepsilon_2(\phi)$ is given by
\begin{align}
\varepsilon_{1}(\phi)-\varepsilon_{2}(\phi)= \frac{(\Delta_0/2) (L_2-L_1)~\text{sin}~\phi}{\left(1+L_1~ \text{sin}(\phi/2) \right)\left(1+L_2~ \text{sin}(\phi/2) \right)}. \label{AE_difference}
\end{align}
This clearly shows a spin-splitting of ABSs and also manifests that the splitting comes from the finite value of $L_2-L_1 \propto (v_1 - v_2) L$. The degeneracies of the Andreev levels at $\phi = 0$ and $\pi$ are protected by the time reversal symmetry~\cite{Padurariu2010,Beri2008}. 
%The split Andreev levels have different spin polarizations, see App.~\ref{app:ABS}. 
%zero current operator matrix element ?

\begin{figure}[!t]
\includegraphics[width=\columnwidth]{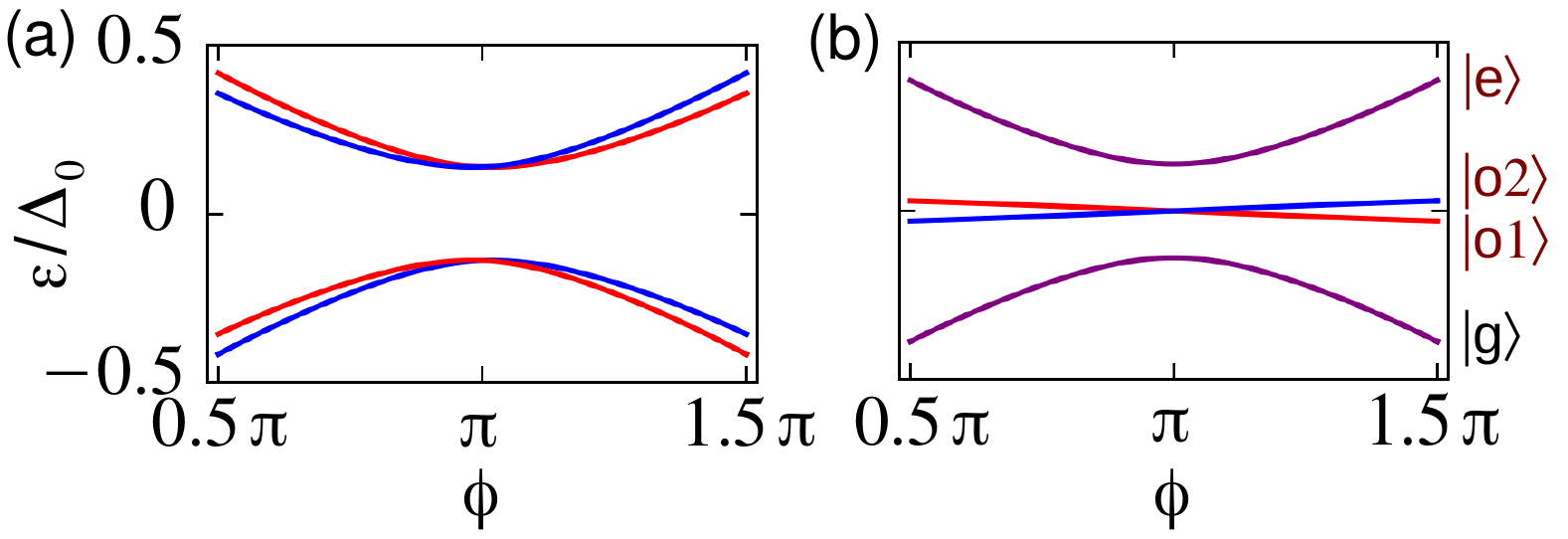}
\caption{Subgap energies of the Josephson junction as a function of the superconducting phase difference $\phi$ without Zeeman field. 
(a) Andreev levels plotted from Eq.~\eqref{Andreevlevels}. The levels colored blue and red are formed by the Andreev reflection processes 
marked by blue and red dashed lines in Fig.~\ref{Fig1:Setup}, respectively. They have spinor structures orthogonal to each other, but 
are coupled through the current operator if Zeeman field $B_x$ is finite. (b) Same plot as in (a), but in the occupation number picture. Two even states, the ground state $|g\rangle$ and excited state $|e\rangle$, and two odd states, $|o1\rangle$ and $|o2\rangle$, are present, where spin splitting between the odd states due to finite values of the Fermi velocity difference and $L$ appears except for $\phi = \pi$. In (a) and (b), we used system parameters $\hbar \alpha = 20$ meV nm, $W = 200$ nm, $L = 300$ nm, $\Delta_0 = 165~\mu$eV, $g$-factor $= 12$, $U_0 = 16.5$ meV nm, $\mu = 0.5$ meV, and $m=0.023~m_e$.
}
\label{Fig2:AE}
\end{figure}

We include the effects of the potential barrier $U_{b}(x)$ which tune the junction transmission and the Zeeman field $H'_Z$ by using perturbation theory. We map $U_{b}(x)$ and $H'_Z$ onto the subspace spanned by the basis $\{\Psi_{1+},\Psi_{1-},\Psi_{2+},\Psi_{2-}\}$, leading to a mapped BdG Hamiltonian $H^{P}_{\text{BdG}}$ as
\begin{align}
H^{P}_{\text{BdG}} = 
\begin{pmatrix}
\varepsilon_1 + \mathbb{B}_{y1} & 0 & \mathbb{B}_x & \mathbb{U} \\
0 & -\varepsilon_1 - \mathbb{B}_{y1} & \mathbb{U}^{*} & \mathbb{B}^{*}_x \\
\mathbb{B}^{*}_x & \mathbb{U} & \varepsilon_2 - \mathbb{B}_{y2} & 0 \\
\mathbb{U}^{*} & \mathbb{B}_x & 0 & -\varepsilon_2 + \mathbb{B}_{y2}
\end{pmatrix}, \label{PBdG}
\end{align}
where $\left( H^{P}_{\text{BdG}}\right)_{j k}$ is computed by
\begin{align}
\left( H^{P}_{\text{BdG}}\right)_{j k} = \int^{\infty}_{-\infty} dx
\Psi^{\dagger}_{j}(x) H^{1D}_{\text{BdG}} \Psi_{k}(x), 
\end{align}
where $j,k\in \{1+,1-,2+,2-\}$. 
The $\mathbb{U}$ term is $U_{b}(x)$ expanded in this basis and is given by 
\begin{align}
\mathbb{U} = -i 2 U_0 e^{i(k_{F_1}+k_{F_2})x_0} 
\sqrt{\frac{\kappa_1 \kappa_2}{N_1 N_2}}~ \text{cos}\left(\frac{\theta_1-\theta_2}{2} \right), \label{PU}
\end{align}
and the Zeeman terms expanded in the basis have the forms
\begin{align}
&\mathbb{B}_{y1(y2)} = \frac{g \mu_B B_y}{2}~ \text{cos}~ \theta_{1(2)}, \label{PBy}\\
&\mathbb{B}_x = i 2 \left(\frac{g \mu_B B_x}{2}\right) \sqrt{\frac{\kappa_1 \kappa_2}{N_1 N_2}} \frac{\kappa_1 + \kappa_2}{(k_{F_1} - k_{F_2})^2}  \nonumber\\
&\hspace{50pt}\times\left( 1+ e^{i(k_{F_1} - k_{F_2}) L}\right) \text{cos}\left(\frac{\theta_1-\theta_2}{2} \right), \label{PBx}
\end{align}
where $\kappa_{1(2)} = (1/(\hbar v_{1(2)})) \sqrt{\Delta^2_0-\varepsilon^2_{1(2)}(\phi)}$ and $N_{1(2)} = 2 (1+ \kappa_{1(2)} L)$. 
In deriving Eq.~\eqref{PBx}, we assumed that $|k_{F_1}-k_{F_2}| \gg |\kappa_1 + \kappa_2|$.
The Hamiltonian $H^{P}_{\text{BdG}}$ is a good approximation provided that $|\mathbb{U}|, |\mathbb{B}_x|, |\mathbb{B}_{y1}|, |\mathbb{B}_{y2}| \ll \Delta_0$ and that $\phi \sim \pi$ where Andreev levels are close to zero energy. 
The $H^{P}_{\text{BdG}}$ reflects the properties of the ABSs $\Psi_{j\lambda}$. 
For the diagonal elements, the $+/-$ sign in front of the terms $\mathbb{B}_{y1}$ (or $\mathbb{B}_{y2}$) indicates the spin polarization direction of the corresponding basis state. As $U_{b}(x)$ is spin-conserving scattering, we have the off-diagonal element $\mathbb{U}$ which couples the basis states of the same spin polarization, i.e., $\Psi_{1-}$ and $\Psi_{2+}$, or $\Psi_{1+}$ and $\Psi_{2-}$, shown in Eqs.~\eqref{Structure_1} and \eqref{Structure_2}. The Zeeman component in the $x$-direction which results in the $\mathbb{B}_x$ element mixes the different spin states, $\Psi_{1\pm}$ and $\Psi_{2\pm}$, but does not mix $\Psi_{j+}$ and $\Psi_{j-}$ (with $j=1,2$) due to the cancellation of contributions from an electron and a hole. Note that the magnitude of $\mathbb{B}_x$ is significantly reduced from its bare value $g \mu_B B_x/2$ by the factor $\sqrt{\kappa_1 \kappa_2} (\kappa_1 + \kappa_2)/(k_{F_1}-k_{F_2})^2$, and oscillates with the length $L$. $H^{P}_{\text{BdG}}$ has two positive Andreev levels, $\varepsilon^{+}_{A1}(\phi)$ and 
$\varepsilon^{+}_{A2}(\phi)$, and two negative Andreev levels, $\varepsilon^{-}_{A1}(\phi)=-\varepsilon^{+}_{A1}(\phi)$ and 
$\varepsilon^{-}_{A2}(\phi)=-\varepsilon^{+}_{A2}(\phi)$: 
\begin{align}
&\varepsilon^{+}_{A1}(\phi) 
= \sqrt{\left(\frac{\varepsilon_1(\phi) + \varepsilon_2(\phi) 
+ \mathbb{B}_{y1} -\mathbb{B}_{y2}}{2} \right)^{2} + |\mathbb{U}|^{2}} \nonumber\\
&\hspace{20pt}- \sqrt{\left(\frac{\varepsilon_1(\phi) - \varepsilon_2(\phi) 
+ \mathbb{B}_{y1} + \mathbb{B}_{y2}}{2} \right)^{2} + |\mathbb{B}_x|^{2}}, \nonumber\\
&\varepsilon^{+}_{A2}(\phi)
= \sqrt{\left(\frac{\varepsilon_1(\phi) + \varepsilon_2(\phi) 
+ \mathbb{B}_{y1} -\mathbb{B}_{y2}}{2} \right)^{2} + |\mathbb{U}|^{2}} \nonumber\\
&\hspace{20pt}+ \sqrt{\left(\frac{\varepsilon_1(\phi) - \varepsilon_2(\phi) 
+ \mathbb{B}_{y1} + \mathbb{B}_{y2}}{2} \right)^{2} + |\mathbb{B}_x|^{2}}.  \label{Andreevlevels}
\end{align}
These Andreev energy levels are plotted in Fig.~\ref{Fig2:AE}(a) in the absence of Zeeman field and for realistic parameters.
The corresponding normalized ABSs are given by 
\begin{align}
\Psi^{+}_{A1}(\phi) &= -\Xi \Psi^{-}_{A1}(\phi) = \frac{1}{\sqrt{N(\phi)}}
\begin{pmatrix}
\tilde{f}(\phi) g(\phi) \\
- f(\phi) \tilde{g}^{*}(\phi) \\
- g(\phi) \tilde{g}^{*}(\phi) \\
f(\phi) \tilde{f}(\phi)
\end{pmatrix}, \nonumber\\
\Psi^{+}_{A2}(\phi) &= \Xi \Psi^{-}_{A2}(\phi) = \frac{1}{\sqrt{N(\phi)}}
\begin{pmatrix}
g(\phi) \tilde{g}(\phi)  \\
f(\phi) \tilde{f}(\phi) \\
\tilde{f}(\phi) g(\phi) \\
f(\phi) \tilde{g}(\phi)
\end{pmatrix}, \label{ABS}
\end{align}
where $\Xi$ is the particle hole symmetry operator, 
%\begin{align}
%\Xi = \begin{pmatrix}
%I & 0 \\
% 0 & -I
%\end{pmatrix} \mathcal{C},
%\end{align}
\begin{align}
\Xi = \begin{pmatrix}
0 & 1 & 0 & 0 \\
1 & 0 & 0 & 0 \\
0 & 0 & 0 & -1 \\
0 & 0 & -1 & 0 
\end{pmatrix} \mathcal{C},
\end{align}
satisfying $\Xi H^{P}_{\text{BdG}}(\phi) \Xi^{-1} = - H^{P}_{\text{BdG}}(\phi)$. 
The components of the ABSs are 
\begin{align}
f(\phi) &= \varepsilon^{+}_{A1}(\phi) +\varepsilon^{+}_{A2}(\phi) - \varepsilon_1(\phi) - \varepsilon_2(\phi) - \mathbb{B}_{y1} +\mathbb{B}_{y2}, \nonumber\\ 
\tilde{f}(\phi) &= -\varepsilon^{+}_{A1}(\phi) +\varepsilon^{+}_{A2}(\phi) - \varepsilon_1(\phi) + \varepsilon_2(\phi) - \mathbb{B}_{y1} - \mathbb{B}_{y2}, \nonumber\\ 
g(\phi) &= 2 \mathbb{U}, \nonumber\\
\tilde{g}(\phi) &= 2 \mathbb{B}_x, \label{ABS_components}
\end{align}
and $N(\phi) = 4 \left[(\varepsilon^{+}_{A2}(\phi))^{2} - (\varepsilon^{+}_{A1}(\phi))^{2}\right] f(\phi)\tilde{f}(\phi)$ is the normalization factor. 
The energy difference between $\varepsilon^{+}_{A1}(\phi)$ and $\varepsilon^{+}_{A2}(\phi)$, 
which corresponds to the splitting of two odd states defined in Eq.~\eqref{Odd_states} below, is given by 
\begin{align}
&|\varepsilon^{+}_{A1}(\phi)-\varepsilon^{+}_{A2}(\phi)| \nonumber\\ 
& \hspace{10pt}= 2  \sqrt{\left(\frac{\varepsilon_1(\phi) - \varepsilon_2(\phi) 
+ \mathbb{B}_{y1} + \mathbb{B}_{y2}}{2} \right)^{2} + |\mathbb{B}_x|^{2}}. \label{Odd_energy}
\end{align} 
We note that it is independent of $\mathbb{U}$ and hence a transmission probability in the normal region.  
These are plotted in Figs.~\ref{Fig2:Result1}(a) and \ref{Fig3:Result2}(a) for different values of $\mu, B_x$, and $B_y$. 
On the other hand, their sum $|\varepsilon^{+}_{A1}(\phi)+\varepsilon^{+}_{A2}(\phi)|$, which is the energy difference between ground and excited states (see Eq.~\eqref{Even_states}), 
\begin{align}
&|\varepsilon^{+}_{A1}(\phi)+\varepsilon^{+}_{A2}(\phi)| \nonumber\\ 
& \hspace{10pt}= 2  \sqrt{\left(\frac{\varepsilon_1(\phi) + \varepsilon_2(\phi) 
+ \mathbb{B}_{y1} -\mathbb{B}_{y2}}{2} \right)^{2} + |\mathbb{U}|^{2}},\label{Even_energy}
\end{align} 
depends on $\mathbb{U}$, but is independent of $\mathbb{B}_x$, as shown in Fig.~\ref{Fig2:Result1}(c). 
Moreover the dependence on $B_y$ is very weak, as shown in Fig.~\ref{Fig3:Result2}(c), in comparison with the dependence of the odd states 
plotted in Fig.~\ref{Fig3:Result2}(a). 
This can be understood by comparing the terms $\mathbb{B}_{y1} + \mathbb{B}_{y2}$ in Eq.~\eqref{Odd_energy} and $\mathbb{B}_{y1} - \mathbb{B}_{y2}$ in Eq.~\eqref{Even_energy} in the limit $|\mu|, |E^{\perp}_{-}| \gg m \alpha^2, \eta$, 
\begin{align}
&\mathbb{B}_{y1} + \mathbb{B}_{y2} \approx -g \mu_B B_y, \nonumber\\
&\mathbb{B}_{y1} - \mathbb{B}_{y2} \approx g \mu_B B_y \frac{\alpha \eta^2 E^{\perp}_{-} \sqrt{2 m \mu}}{\left[ (E^{\perp}_{-})^2 -2 \alpha^2 m \mu \right]^2}, \label{Approx_By}
\end{align} 
where we used Eq.~\eqref{Approx_theta}. Therefore, this implies that $|\mathbb{B}_{y1} + \mathbb{B}_{y2}| \gg |\mathbb{B}_{y1} - \mathbb{B}_{y2}|$ leads 
to the strong (weak) dependence of the odd (even) states on $B_y$. 
However, it is found that changing $\mu$ changes both $|\varepsilon^{+}_{A1}-\varepsilon^{+}_{A2}|$ and $|\varepsilon^{+}_{A1}+\varepsilon^{+}_{A2}|$, as shown in Figs.~\ref{Fig2:Result1}(a), (c) and \ref{Fig3:Result2}(a), (c).
The different dependencies of the even and odd states on the system parameters allow us to control $|\varepsilon^{+}_{A1}(\phi)-\varepsilon^{+}_{A2}(\phi)|$ independently by changing $B_x$ or $B_y$ without changing $|\varepsilon^{+}_{A1}(\phi)+\varepsilon^{+}_{A2}(\phi)|$. This is one of our main results. 
%We note that in the limits of $\eta \rightarrow 0$ and $B_x \rightarrow 0$, the 
%Andreev bound states have definite spin polarizations such that their spin 
%polarization directions for the electron sector are 
%\begin{align}
%\Psi^{+}_{A1}(\phi), \Psi^{-}_{A2}(\phi)\big|_{\text{electron}} \propto \chi_
%{\downarrow}, \nonumber\\
%\Psi^{-}_{A1}(\phi), \Psi^{+}_{A2}(\phi)\big|_{\text{electron}} \propto \chi_
%{\uparrow}.
%\end{align} 

\section{Current operator}\label{Current}

To describe the microwave response of the nanowire Josephson junction, 
we calculate the current operator matrix, whose off-diagonal elements determine the transitions induced by the coupling to the external radiation, in the subspace of the low-energy ABSs given in Eq.~\eqref{ABS} and analyze their dependence on the system parameters.  
In the subgap energy region, there are two even states, ground state $|g\rangle$ with an energy $(\varepsilon^{-}_{A1}+\varepsilon^{-}_{A2})/2$ and 
excited state $|e\rangle$ with an energy $(\varepsilon^{+}_{A1}+\varepsilon^{+}_{A2})/2$. The states are defined by  
\begin{align}
\gamma_{A1+}|g\rangle = \gamma_{A2+}|g\rangle =0,\hspace{20pt}
|e\rangle = \gamma^{\dagger}_{A1+} \gamma^{\dagger}_{A2+}|g\rangle,  \label{Even_states} 
\end{align}
where $\gamma_{A1\pm(A2\pm)} = \int dx (\Psi^{\pm}_{A1(A2)}(x))^{\dagger} \Phi (x)$, with the Nambu field operator $\Phi (x)$, are the Bogoliubov operators. 
By adding or removing a single quasiparticle from the even states, we have 
two odd states $|o1\rangle$ and $|o2\rangle$, 
\begin{align}
|o1\rangle = \gamma^{\dagger}_{A1+}|g\rangle,\hspace{20pt} 
|o2\rangle = \gamma^{\dagger}_{A2+}|g\rangle, \label{Odd_states}
\end{align} 
and their energies are $(\varepsilon^{+}_{A1}+\varepsilon^{-}_{A2})/2$ and $(\varepsilon^{-}_{A1}+\varepsilon^{+}_{A2})/2$, respectively. 
Fig.~\ref{Fig2:AE}(b) shows the plot of these energies of the even and odd states in the case of zero Zeeman field. 
The particle hole symmetry of the ABSs given in Eq.~\eqref{ABS} implies the relations 
\begin{align}
\gamma^{\dagger}_{A1+} = -\gamma_{A1-}, \hspace{20pt} 
\gamma^{\dagger}_{A2+} = \gamma_{A2-}.
\end{align}

\begin{figure}[!t]
\centering
\includegraphics[width=\columnwidth]{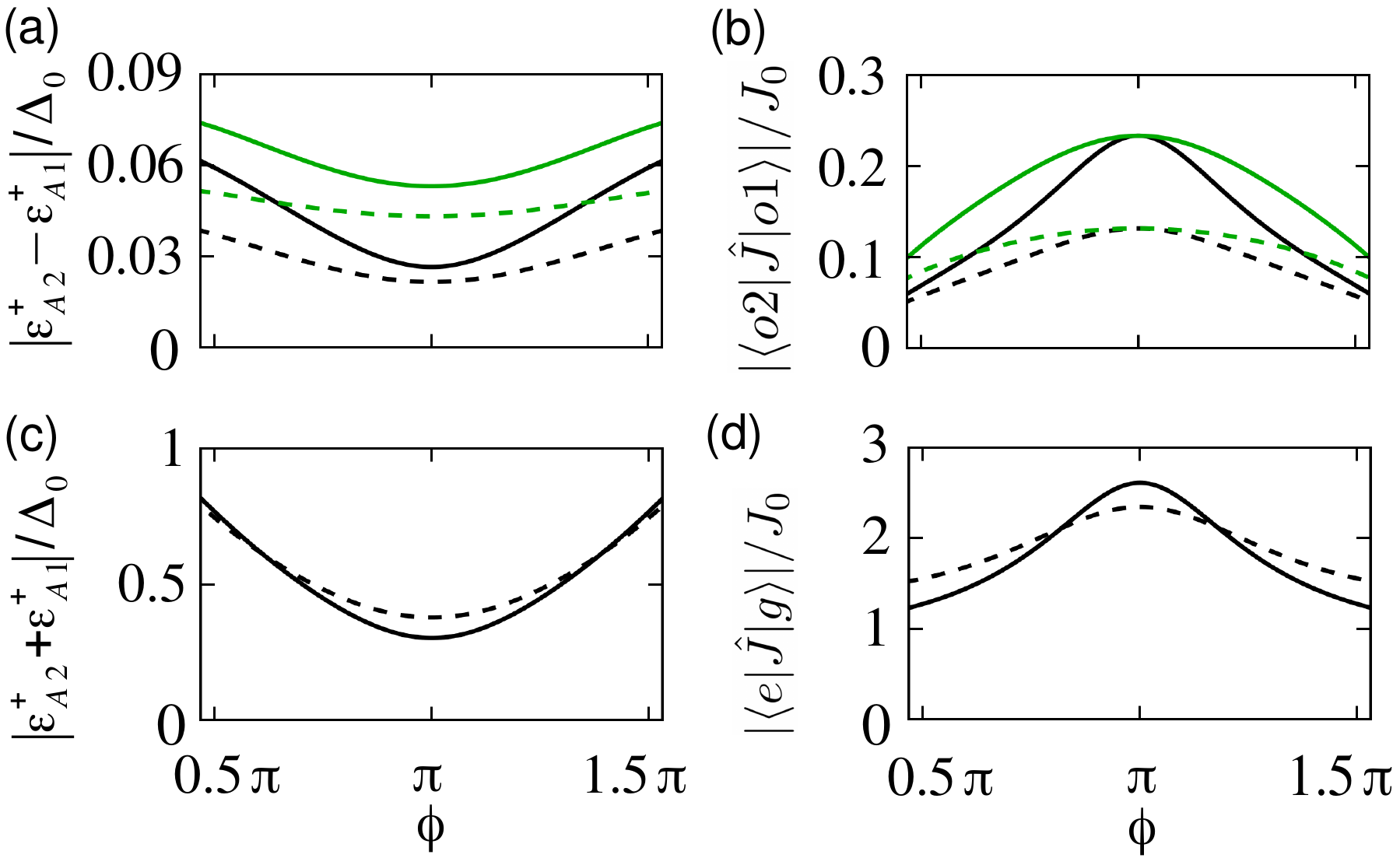}
\caption{Excitation spectra and matrix elements of the current operator (in units of $J_0 = e \Delta_0/h$) 
as a function of $\phi$ at $B_y =0$ for odd (a, b) and even (c, d) transitions [see Eqs.~\eqref{Andreevlevels}, \eqref{EvenParity}, and \eqref{OddParity}]. We plot for different values of $\mu$ and $B_x$; $\mu = 0.51$ meV and $B_x = 50$ mT (black solid lines), $0.41$ meV and $50$ mT (black dashed), $0.51$ meV and $100$ mT (green solid), and $0.41$ meV and $100$ mT (green dashed).
The other system parameters are the same as in Fig.~\ref{Fig2:AE}. Contrary to 
the results (a) and (b) for the odd states which depend on both $\mu$ and $B_x$, the results (c) and (d) for the even states are 
independent of the value of $B_x$. The heights of the peaks at $\phi = \pi$ shown in (b) and (d) depend on $\mu$ but are independent of $B_x$.  
}
\label{Fig2:Result1}
\end{figure}

\begin{figure}[!t]
\centering
\includegraphics[width=\columnwidth]{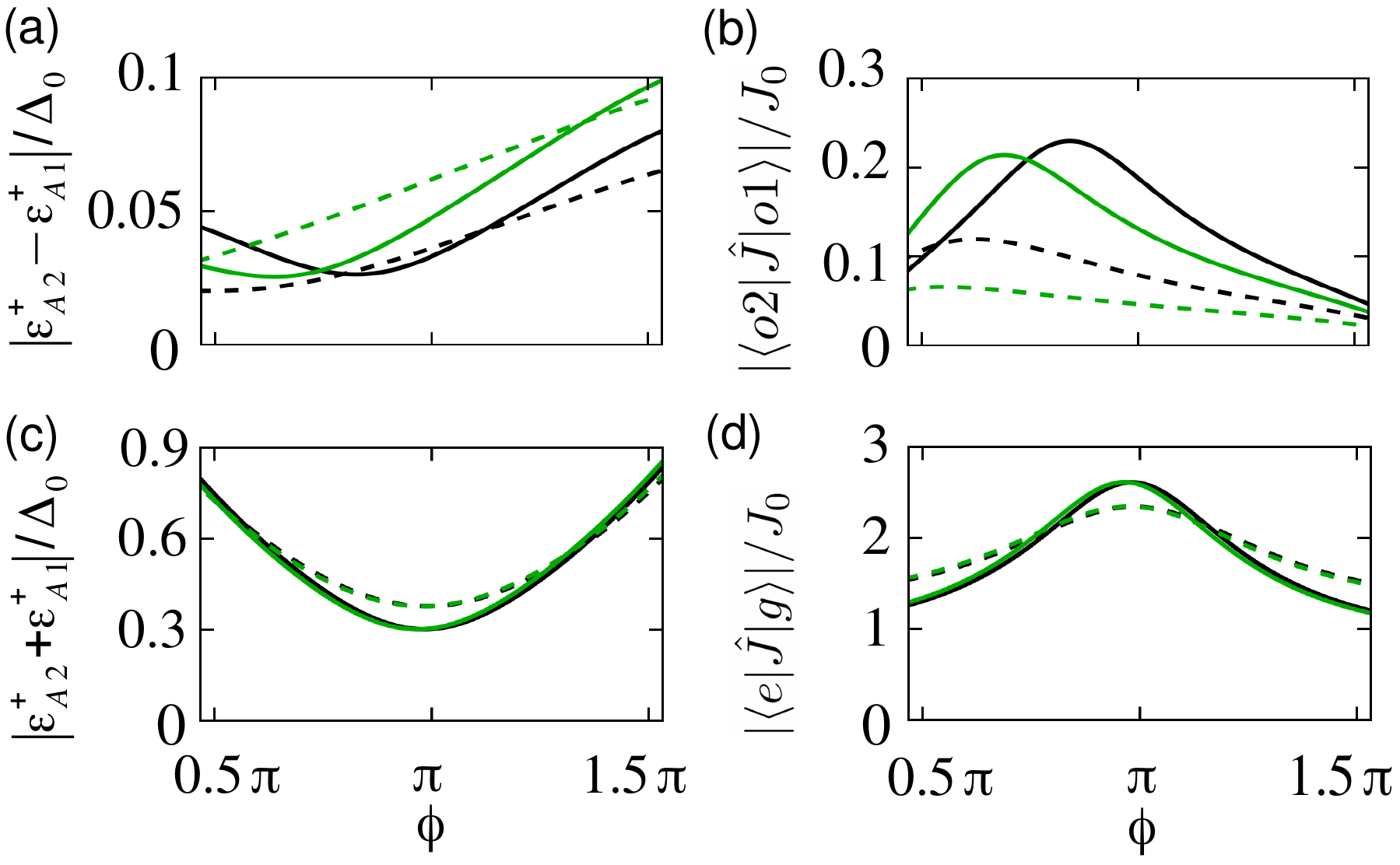}
\caption{(a)-(d) Same plots as in Fig.~\ref{Fig2:Result1}, but for different values of $B_y$; $\mu = 0.51$ meV and $B_y = 10$ mT (black solid lines), $0.41$ meV and $10$ mT (black dashed), $0.51$ meV and $20$ mT (green solid), 
and $0.41$ meV and $20$ mT (green dashed). Here $B_x = 50$ mT is used.
%and other system parameters are the same as in Fig.~\ref{Fig2:Result1}. 
The excitation spectrum $|\varepsilon^{+}_{A2}+\varepsilon^{+}_{A1}|$ and the $|\langle e |\hat{J}|g \rangle|$ for the even states are weakly dependent on $B_y$ compared to the dependence for the odd states.
}
\label{Fig3:Result2}
\end{figure}

The current operator for the BdG Hamiltonian $H^{1D}_{\text{BdG}}$ in Eq.~\eqref{1DBdG} is 
\begin{align}
\hat{J} = \sum_{m,n} J_{m,n} \hat{\gamma}^{\dagger}_{m}\hat{\gamma}_{n}, 
\end{align}
where $m,n \in \{A1+,A1-,A2+,A2-\}$. $J_{m,n}$ are the matrix elements of the current operator, and are obtained from 
the ABSs in Eq.~\eqref{ABS}~\cite{Olivares2014}. The diagonal matrix elements determine the supercurrent
carried by even and odd states. In the ground and excited states, these are 
\begin{align}
\langle g | \hat{J} | g \rangle &= - \langle e | \hat{J} | e \rangle \nonumber\\
&= \sum_{m=A1-,A2-} J_{m,m} \langle g |\gamma^{\dagger}_{m} \gamma_{m}  | g \rangle  \nonumber\\ 
&= J_{A1-,A1-} +  J_{A2-,A2-}, 
\end{align} 
and in the odd states, 
\begin{align}
\langle o1 | \hat{J} | o1 \rangle &= -  \langle o2 | \hat{J} | o2 \rangle \nonumber\\ 
&=\langle g | \gamma_{A1+}\hat{J} \gamma^{\dagger}_{A1+} | g \rangle \nonumber\\
&=  J_{A1+,A1+},
\end{align}
where the matrix elements are given by
\begin{align}
J_{A1+,A1+} = -J_{A1-,A1-} =  \frac{-e}{\hbar}\frac{\partial \varepsilon^{+}_{A1}(\phi)}{\partial \phi}, \nonumber\\
J_{A2+,A2+} = -J_{A2-,A2-} = \frac{-e}{\hbar}\frac{\partial \varepsilon^{+}_{A2}(\phi)}{\partial \phi}.
\end{align}
The current matrix element between the ground and excited states $\langle e| \hat{J} |g \rangle$ is 
\begin{align}
\langle e| \hat{J} |g \rangle &= \langle g| \gamma_{A2+} \gamma_{A1+}\hat{J} |g \rangle \nonumber\\
&=  J_{A1+,A2-} +  J_{A2+,A1-},  \label{EvenParity}
\end{align}
and the element between the odd states is 
\begin{align}
\langle o2| \hat{J} |o1 \rangle &= \langle g| \gamma_{A2+} \hat{J} \gamma^{\dagger}_{A1+} |g \rangle \nonumber\\
&=  J_{A2+,A1+} +  J_{A1-,A2-}, \label{OddParity}
\end{align}
where 
\begin{align}
J_{A1+,A2-} &= J_{A2+,A1-} = \left(J_{A2-,A1+}\right)^{*} = \left(J_{A1-,A2+}\right)^{*} \nonumber\\
&= \frac{-e}{2 \hbar} \frac{g^{*}(\phi)}{\varepsilon^{+}_{A2}(\phi) + \varepsilon^{+}_{A1}(\phi)} \frac{\partial (\varepsilon_{1}(\phi) + \varepsilon_{2}(\phi))}{\partial \phi}. \label{EvenParity_2}
\end{align}
and 
\begin{align}
J_{A1+,A2+} &= J_{A2-,A1-} = \left(J_{A2+,A1+}\right)^{*} = \left(J_{A1-,A2-}\right)^{*} \nonumber\\
&= \frac{-e}{2 \hbar} \frac{\tilde{g}(\phi)}{\varepsilon^{+}_{A2}(\phi) - \varepsilon^{+}_{A1}(\phi)} \frac{\partial (\varepsilon_{1}(\phi) - \varepsilon_{2}(\phi))}{\partial \phi}. \label{OddParity_2}
\end{align}
The remaining matrix elements $J_{A1+,A1-} = (J_{A1-,A1+})^{*}$ and $J_{A2+,A2-} = (J_{A2-,A2+})^{*}$ are zero as $\gamma^{\dagger}_{Aj+} \gamma_{Aj-} = (-1)^{j}(\gamma_{Aj-})^2 = 0$, where $j=1,2$. 
%The results for $|\langle o2| \hat{J} |o1 \rangle|$ and $|\langle e| \hat{J} |g \rangle|$ 
%are plotted in Figs.~\ref{Fig2:Result1} and \ref{Fig3:Result2}.
Below, we discuss the dependence of $\langle o2| \hat{J} |o1 \rangle$ and $\langle e| \hat{J} |g \rangle$ on tunable system parameters, like $B_{x,y}$, $v_j$, $L$, and $\mu$.

Before discussing in detail the dependence, we examine the case of $\eta = 0$ and $L \rightarrow 0$ in order to check the consistency of our perturbative results with previous theoretical~\cite{Zazunov2003,Kos2013,Olivares2014,Ivanov1999,Desposito2001} as well as experimental~\cite{Janvier2015,Zgirski2011} studies on Josephson junctions in the short-junction limit. 
As there is no transverse-subband mixing in this case, we have $v_1 = v_2$ and 
\begin{align}
\varepsilon_1(\phi) = \varepsilon_2(\phi) = \Delta_0 \text{cos} \frac{\phi}{2},\hspace{15pt} 
\mathbb{U} = -i \frac{U_0 \Delta_0}{\hbar v_1} \left|\text{sin}\frac{\phi}{2} \right|. 
\end{align} 
Then from Eq.~\eqref{OddParity_2} we see that $\langle o2| \hat{J} |o1 \rangle = 0$,   
regardless of $B_{x,y}$ and $\mu$. On the other hand, we find for the even states that 
\begin{align}
\langle e| \hat{J} |g \rangle\big|_{\eta, L =0} = \frac{-2e}{\hbar} \frac{\mathbb{U}^{*}}{\sqrt{\varepsilon_{1}^{2}(\phi)+|\mathbb{U}|^2}}\frac{\partial \varepsilon_{1} (\phi)}{\partial \phi}. 
\end{align}  
If we further assume that the Zeeman field is absent, it can be expressed as 
\begin{align}
\langle e| \hat{J} |g \rangle\big|_{\eta, L, B_x, B_y =0} = -i \frac{e}{\hbar} \frac{\Delta^{2}_0 \sqrt{1-T}}{\varepsilon_A(\phi)} \text{sin}^{2}\frac{\phi}{2},
\end{align} 
where $T = 1-|U_0/(\hbar v_{1})|^2$ is the transmission probability in the normal region in our weak scattering limit 
and $\varepsilon_A(\phi) = \Delta_0 \sqrt{1-T~ \text{sin}^2(\phi/2)}$. This result is consistent with the previous results~\cite{Janvier2015,Desposito2001} in the limit of perfect transmission. As already known, this even transition matrix element is finite 
even in the absence of effects of Rashba spin-orbit, Zeeman, and multichannel structure.   

With finite $\eta$ and $L$, we analyze the matrix elements between the even and odd states by considering their dependence on $\mathbb{U}$, $\mathbb{B}_x$, and $\mathbb{B}_{y1,y2}$. From Eqs.~\eqref{EvenParity} and \eqref{OddParity}, we get 
\begin{align}
\langle e| \hat{J} |g \rangle &\propto  \frac{\mathbb{U}^{*}}{\varepsilon^{+}_{A2}(\phi) + \varepsilon^{+}_{A1}(\phi)},  \nonumber\\
\langle o2| \hat{J} |o1 \rangle &\propto \frac{\mathbb{B}_x}{\varepsilon^{+}_{A2}(\phi) - \varepsilon^{+}_{A1}(\phi)}.
\end{align}
This even-(odd-) state matrix element follows the same dependence of its energy $|\varepsilon^{+}_{A2} + \varepsilon^{+}_{A1}| (|\varepsilon^{+}_{A2} - \varepsilon^{+}_{A1}|)$ on the system parameters which we discussed above. Specifically, varying the parameter $\mathbb{U}(\mathbb{B}_x)$ changes the element $\langle e| \hat{J} |g \rangle (\langle o2| \hat{J} |o1 \rangle)$ while the other element $\langle o2| \hat{J} |o1 \rangle (\langle e| \hat{J} |g \rangle)$ remains unchanged, as clearly shown in Fig.~\ref{Fig2:Result1}(b) and (d) in which these elements are plotted for different values of $B_x$. 
Also, due to the dependence of the energies on $B_y$ that is described by Eq.~\eqref{Approx_By}, 
the $|\langle o2| \hat{J} |o1 \rangle|$ term shows a significant change with $B_y$ (Fig.~\ref{Fig3:Result2}(b)), but 
there is a small change of $|\langle e| \hat{J} |g \rangle|$ on $B_y$ (Fig.~\ref{Fig3:Result2}(d)). 

We consider the matrix elements at $\phi = \pi$ for further detailed analysis. The $\langle o2| \hat{J} |o1 \rangle$ term at $\phi = \pi$ is obtained from Eqs.~\eqref{AE_difference}, \eqref{Andreevlevels}, and \eqref{ABS_components}: 
\begin{align}
\langle o2| \hat{J} |o1 \rangle\big|_{\phi=\pi} &= \frac{-e\Delta_0}{2\hbar}\frac{\mathbb{B}_x}{\sqrt{\left(\mathbb{B}_{y1}+\mathbb{B}_{y2}\right)^2/4 + |\mathbb{B}_x|^2}} \nonumber\\
&\hspace{30pt}\times \frac{L_1 - L_2}{(1+L_1)(1+L_2)},
\end{align}  
where $L_j = \Delta_0 L/(\hbar v_j)$. As this element is proportional to $\mathbb{B}_x (L_1-L_2)$, 
the finite values of $B_x$, $L$, and $|v_1-v_2|$ are required in order to be nonzero. 
When we assume that $B_y=0$, its magnitude can be further simplified as 
\begin{align}
&\left|\langle o2| \hat{J} |o1 \rangle\right|\big|_{\phi=\pi, B_{y} =0} = 
\frac{e\Delta_0}{2\hbar}\frac{L_1 - L_2}{(1+L_1)(1+L_2)} \nonumber\\
&\hspace{20pt} = \frac{e\Delta_0}{2\hbar}\frac{L_1 - L_2}{\left[1+(L_1+L_2)/2\right]^2} + \mathcal{O}((L_1-L_2)^3), \label{Odd_transition_sim}
\end{align} 
which is independent of both $\mathbb{B}_x$ and $\mathbb{U}$, 
except for a singular value $\mathbb{B}_x = 0$ where $\langle o2| \hat{J} |o1 \rangle = 0$. 
The independence on $\mathbb{B}_x$ is shown in Fig.~\ref{Fig2:Result1}(b) in which the peak heights of 
$\langle o2| \hat{J} |o1 \rangle$ at $\phi=\pi$ remain unchanged for different values of $B_x$. 
For the dependence on $L$, Eq.~\eqref{Odd_transition_sim} has its maximum value at $L=L_c$ where 
\begin{align}
L_c= \frac{2 \hbar}{\Delta_0}\left(\frac{1}{v_1} + \frac{1}{v_2} \right)^{-1},
\end{align}
in the limit $|L_1 - L_2|\ll 1$. 
A word of caution should be said regarding the validity of this $L_c$ estimation, which is of the order of the coherence length $\hbar v_j/\Delta_0$. 
The energy-phase relation $\varepsilon_j(\phi)$ in Eq.~\eqref{EnergyPhaseRelation} 
is valid when either $\Delta_0 L / (\hbar v_j) \ll 1$ or $\varepsilon \ll \Delta_0$ is fulfilled. Therefore, the $L_c$ might be qualitatively correct as 
$\varepsilon_{j}(\phi)=0 \ll \Delta_0$ at $\phi=\pi$. 
The $\langle e| \hat{J} |g \rangle$ matrix element at $\phi=\pi$, which is obtained by
\begin{align}
\langle e| \hat{J} |g \rangle\big|_{\phi=\pi} &= \frac{e\Delta_0}{2\hbar}\frac{\mathbb{U}^{*}}{\sqrt{\left(\mathbb{B}_{y1}-\mathbb{B}_{y2}\right)^2/4 + |\mathbb{U}|^2}} \nonumber\\
&\hspace{30pt}\times \frac{2 + L_1 + L_2}{(1+L_1)(1+L_2)},
\end{align}  
is independent of $\mathbb{B}_x$ but depends on $\mathbb{U}$ which is associated with the transmission probability in the normal region. 
However, similar to the case of $\langle o2| \hat{J} |o1 \rangle$, if $B_{y}=0$,
the magnitude of this element does not depend on both $\mathbb{B}_x$ and $\mathbb{U}$ as 
\begin{align}
&\left|\langle e| \hat{J} |g \rangle\right|\big|_{\phi=\pi, B_{y} =0} = 
\frac{e\Delta_0}{2\hbar}\frac{2 + L_1 + L_2}{(1+L_1)(1+L_2)} \nonumber\\
&\hspace{20pt} = \frac{e\Delta_0}{\hbar}\frac{1}{1+(L_1+L_2)/2} + \mathcal{O}((L_1-L_2)^2),
\end{align} 
except for a singularity of $\mathbb{U} =0$ where $\langle e| \hat{J} |g \rangle =0$. Note also that it decreases as $L$ increases. 

In the above calculation, we have neglected the orbital effect of a magnetic field $B_x$, which would lead to a longitudinal magnetic flux $\Phi$ piercing our cylindrical nanowire. In App.~\ref{app:Orbital_effect}, we show that there is no first order correction to 
the dispersion relation in Eq.~\eqref{Dispersion}, and the leading order correction is of second order in $\Phi$. 
Therefore the above results for the ABSs and the matrix elements might be still valid 
up to first order in $B_x$ with respect to the orbital effect.
  
%\begin{align}
%J_{A1+,A2+} = \frac{-e\Delta_0}{\hbar}\frac{\mathbb{B}_x}{\varepsilon^{+}_{A2}(\pi)-\varepsilon^{+}_{A1}(\pi)} 
%\frac{L_1 - L_2}{(1+L_1)(1+L_2)}
%\end{align}
%
%\begin{align}
%J_{A1+,A2-} = \frac{-e\Delta_0}{\hbar}\frac{\mathbb{U}^{*}}{\varepsilon^{+}_{A2}(\pi)+\varepsilon^{+}_{A1}(\pi)} 
%\frac{L_1 + L_2}{(1+L_1)(1+L_2)}
%\end{align} 

\section{Experimental observation of odd transitions}\label{Exp}

We now briefly discuss the feasibility of observing the odd transitions in an actual experiment.  
We consider an experimental setup where our nanowire Josephson junction is embedded in a superconducting ring which is inductively coupled to a microwave resonator. A similar setup for an superconducting atomic contact was used in Ref.~\cite{Janvier2015}. 
In the dispersive limit (i.e. far from resonance), the visibility of the transition will be determined by the cavity pull $\chi$ fixed by the coupling to the nanowire and which can be written for the case of odd transitions as 
\begin{align}
\chi_{odd} \propto \frac{|\langle o2| \hat{J} |o1 \rangle|^2}{\omega_R - \omega_A},
\end{align} 
where $\hbar \omega_A= |\varepsilon^+_{A1}-\varepsilon^+_{A2}|$ is the Andreev energy level and $\omega_R$ is the resonator frequency. The proportionality constant depends on the mutual inductance and the impedance of the resonator which can be assumed to be of the same order as in Ref.~\cite{Janvier2015}. One stringent condition for the direct detection of the odd transitions is 
\begin{align}
\chi_{odd} > \Delta \omega = \frac{\omega_R}{Q},
\end{align} 
which means that the shift of the resonance frequency set by $\chi_{odd}$ has to be larger than the width of the resonance $\Delta \omega$, which in terms of the
resonator quality factor $Q$ is $\sim \omega_R/Q$. We take as a reference
the typical values of $\chi \sim 3$ MHz 
in the experiments of Ref.~\cite{Janvier2015} where even transitions were observed, i.e. $\chi_{even}\sim 3$ MHz. If we assume similar conditions so that the proportionality constant is the same for both even and odd transitions, we estimate 
$\chi_{odd}$ as
\begin{align}
\chi_{odd} \sim \chi_{even} \left| \frac{\langle o2| \hat{J} |o1 \rangle}{\langle e| \hat{J} |g \rangle}\right|^2 \sim 0.03 \text{MHz},
\end{align} 
where we assume that the Andreev energy levels are much smaller than 
$\omega_R$ and that $|\langle o2| \hat{J} |o1 \rangle/\langle e| \hat{J} |g \rangle| \sim 0.1$ around $\phi = \pi$ from the results shown in Fig.~\ref{Fig2:Result1}. Therefore, if we assume 
$\omega_R \sim 2-10$ GHz, the condition for the quality factor to observe the odd transitions is given by 
\begin{align}
Q > \frac{\omega_R}{\chi_{odd}} \sim 0.6 \times 10^{5} - 0.3 \times 10^{6},
\end{align} 
which is challenging, but still within the present technological capabilities. It should be also noticed that this high $Q$ requirement could be relaxed provided that
a larger inductive coupling between the nanowire junction and the resonator is
achieved or by working with a larger number of photons in the resonator than in Ref.~\cite{Janvier2015}. 

Another approach would be provided by using an indirect detection technique like the {\it shelving} method, which is well known in atomic physics~\cite{Dehmelt1975,Bergquist1986,Sauter1986} but their extension to circuit QED like experiments could be explored~\cite{Devoret}.

\section{Concluding remarks}\label{Conclusion}

We have analyzed the ABSs and the current matrix elements in multichannel nanowire Josephson junctions. We found analytical expressions for the Andreev energy levels and the matrix elements including the effects of a Zeeman field and a potential barrier by using pertubation theory, and investigated their dependence on the system parameters. We have shown that the multichannel structure of the nanowire, in combination with the Rashba spin-orbit interaction, plays a fundamental role in breaking the degeneracy between opposite spin ABSs in the absence of Zeeman field and gives rise to finite matrix elements for transitions between the odd states in the presence of a small Zeeman field. In particular, the energy difference and the matrix elements between the odd states are found to have strong dependence on the field, while those between the even states remain almost unchanged. 
Contrary to the Zeeman effect, the barrier determining the transmission probability in the normal region only affects to the even transitions without affecting the odd transitions. Regarding the dependence of the junction length $L$, there exists a length scale $L_c$ at which the odd transition matrix elements have their maximum, while the corresponding ones for even transitions decrease monotonically with the length. 
%Both even and odd transitions depend on the chemical potential. 
Our results may provide a way to selectively control the even and odd transitions by tuning the system parameters, and could be used
to guide the experiments in the realization of an Andreev spin qubit. 

{\it Note added:} During the process of writing this manuscript we become aware of
a related work by van Heck, V\"{a}yrynen, and Glazman~\cite{Heck2017}, addressing 
the effect of Zeeman and spin-orbit coupling in the properties of Andreev states in semiconducting nanowire junctions. 
We point out that these two works correspond to different regimes, ours being in 
the regime of multichannel and small Zeeman field, and 
the regime of Ref.~\cite{Heck2017} in the single-channel with a wide range of Zeeman field.  
%There are,
%however, important differences between both works: on the one hand that
%work does not take into account the multichannel structure of the wires. On the
%other hand their work is not restricted to the low Zeeman regime.  

\acknowledgments  
We thank B. Braunecker, M. Devoret, M. Goffman, H. Pothier, L. Tosi and C. Urbina for useful discussions.  
This work has been supported by the Spanish MINECO through Grant No.~FIS2014-55486-P
and through the ``Mar\'{\i}a de Maeztu'' Programme for Units of Excellence in R\&D (MDM-2014-0377).

\appendix

\section{Calculation details for Andreev bound states}\label{app:ABS}

In this appendix, we provide the explicit expressions for the Andreev eigenstates $\Psi_{j\lambda}(x)$ 
which are used as the basis for the mapped BdG Hamiltonian given in Eq.~\eqref{PBdG}.
We solve the BdG equations of Eq.~\eqref{1DBdG} in the main text with $U_{b} = 0$ and $H'_{Z} =0$, 
\begin{align}
&H^{1D}_{\text{BdG}}~ \Psi(x) = \varepsilon~ \Psi(x),  \nonumber\\
&H^{1D}_{\text{BdG}} = \left(H'_0-\mu\right) \tau_z + H'_R \tau_z + H_S,
\end{align}
where  $\Psi(x) = (\psi^{e}(x),\psi^{h}(x))^{T}$. 
We consider the chemical potential $\mu$ close to but below the bottom of the second transverse subbands 
$\mu \leq E^{\perp}_2$ that two right and two left moving electron (or hole) waves are present at the Fermi energy in the normal region of the nanowire. 
Next we linearize the dispersion relation around the chemical potential, as shown in Eq.~\eqref{LinearizedDispersion},
\begin{align}
E^{e(h)}_{R,j} &= \mu \pm \hbar v_j \left( k^{e(h)}_{xj} -k_{F_j}\right), \nonumber\\
E^{e(h)}_{L,j} &= \mu \mp \hbar v_j \left( k^{e(h)}_{xj} + k_{F_j}\right),
\end{align}
where $k^{e(h)}_{xj}$ are wave vectors of electrons (holes) at energy $\mu + \varepsilon$ ($\mu - \varepsilon$), and 
$k_{F_j}$ are Fermi wave vectors of electrons shown in Fig.~\ref{Fig1:Setup}. 
Here we assume that perfect Andreev reflection happens at the interface between the normal and superconducting regions, 
meaning that there is no normal or Andreev reflection between the bands except for the electron-hole conversion within 
the linearized band structure, $E^{e}_{R(L),j} \rightleftharpoons E^{h}_{R(L),j}$. 
We further assume that the spinor parts of the wave functions, composed of spin and transverse degree of freedom ($n=1,2$), 
do not change significantly within the subgap energy regime $|\varepsilon| < \Delta_0$. 
This assumption is a good approximation for a large separation between the transverse subbands compared to the induced superconducting gap 
$E^{\perp}_{-} = (E^{\perp}_1 - E^{\perp}_2)/2 \gg \Delta_0$.  

We calculate $\Psi_{j+}(x)$ with $j=1,2$ which are formed by a superposition of the left moving electrons 
and the right moving Andreev reflected holes, 
\begin{align}
\Psi_{j+}(x) = a_{j}(x)~ \chi^{e}_{j,+}(k_{F_j}) + b_{j}(x)~ \chi^{h}_{j,+}(k_{F_j}), \label{app:Eigenstates}
\end{align} 
where $\chi^{e}_{j,+}$ and $\chi^{h}_{j,+}$ are the spinor parts of the states
\begin{align}
 \chi^{e}_{1,+}(k_{F_1}) &= \chi^{h}_{1,+}(k_{F_1}) = 
 \begin{pmatrix}
 0\\
 \text{sin}~(\theta_1(k_{F_1})/2)\\
 \text{cos}~(\theta_1(k_{F_1})/2)\\
 0
 \end{pmatrix}, \nonumber\\
 \chi^{e}_{2,+}(k_{F_2}) &= \chi^{h}_{2,+}(k_{F_2}) = 
 \begin{pmatrix}
 \text{sin}~(\theta_2(k_{F_2})/2)\\
 0\\
 0\\
 -\text{cos}~(\theta_2(k_{F_2})/2)
 \end{pmatrix},
\end{align}  
where 
\begin{align}
\theta_1(k_{F1}) &= \text{arccos}\left[\frac{E^{\perp}_{-} - \alpha \hbar k_{F_1}}{\sqrt{\left( E^{\perp}_{-} - \alpha \hbar k_{F_1}\right)^2 + \eta^2}}\right], \nonumber\\
\theta_2(k_{F2}) &= \text{arccos}\left[\frac{E^{\perp}_{-} + \alpha \hbar k_{F_2}}{\sqrt{\left( E^{\perp}_{-} + \alpha \hbar k_{F_2}\right)^2 + \eta^2}}\right].
\end{align}
Here we used the approximation $\theta_j(k^{e(h)}_{xj})\approx \theta_j(k_{Fj})$ based on the above mentioned assumption that 
the spinor do not change much in the subgap energy range. 
The coefficients $a_{j}(x)$ and $b_{j}(x)$ in Eq.~\eqref{app:Eigenstates} are evaluated by solving the following equation, 
\begin{align}
\begin{pmatrix}
\hbar v_j \left(i \partial_x - k_{F_j}\right) & \Delta(x) e^{i \phi(x)} \\
\Delta(x) e^{-i \phi(x)} & \hbar v_j \left(-i \partial_x + k_{F_j}\right)
\end{pmatrix} 
\begin{pmatrix} 
a_{j}\\
b_{j}
\end{pmatrix} = 
\varepsilon 
\begin{pmatrix} 
a_{j}\\
b_{j}
\end{pmatrix}, \nonumber
\end{align}
where 
\begin{align}
(\Delta(x), \phi(x)) = 
\begin{cases}
(\Delta_0, \phi_L) & \text{for $x < 0$}, \\
(0,0) & \text{for $0 \leq x \leq L$}, \\
(\Delta_0, \phi_R) & \text{for $x > L$}.
\end{cases}
\end{align}
By matching wave functions at the interfaces, we obtain normalized ABSs 
\begin{widetext}
\begin{align}
\Psi_{j+}(x) = \begin{cases}
\sqrt{\frac{\kappa_j}{N_j}} e^{(-i k_{F_j}+\kappa_j) x} 
\begin{pmatrix}
e^{i \phi_L/2}~ \chi^{e}_{j,+}(k_{F_j})\\
e^{-i \phi_L/2}~\beta_j~ \chi^{h}_{j,+}(k_{F_j})
\end{pmatrix} & \text{for  $x<0$}, \\
\sqrt{\frac{\kappa_j}{N_j}} 
\begin{pmatrix}
e^{i \phi_L/2 - i k^{e}_{xj} x}~ \chi^{e}_{j,+}(k_{F_j})\\
e^{-i \phi_L/2 - i k^{h}_{xj} x}~ \beta_j~ \chi^{h}_{j,+}(k_{F_j})
\end{pmatrix} & \text{for $0 \leq x \leq L$}, \\
\sqrt{\frac{\kappa_j}{N_j}} e^{(-i k_{F_j}-\kappa_j) (x-L)} 
e^{-i(\phi_R - \phi_L)/2-i k^{e}_{xj} L}~\beta^{*}_j~
\begin{pmatrix}
e^{i \phi_R/2}~\beta_j~ \chi^{e}_{j,+}(k_{F_j})\\
e^{-i \phi_R/2}~  \chi^{h}_{j,+}(k_{F_j})
\end{pmatrix} &\text{for $L < x$},
\end{cases}
\end{align} 
\end{widetext}
where $\beta_{j} = \varepsilon_{j}(\phi)/\Delta_0 - i \sqrt{1-(\varepsilon_{j}(\phi)/\Delta_0)^2}$, 
$\kappa_j = (1/(\hbar v_j)) \sqrt{\Delta^{2}_0 - \varepsilon^2_{j}(\phi)}$ are the imaginary parts of the momenta related to the exponential decay of wave functions in the superconducting regions, and $N_j = 2 (1 + \kappa_j L)$ are normalization constants.

In a similar way, we calculate $\Psi_{j-}(x)$ expressed as 
\begin{align}
\Psi_{j-}(x) = c_{j}(x)~\chi^{e}_{j,-}(k_{F_j}) + d_{j}(x)~ \chi^{h}_{j,-}(k_{F_j}).
\end{align} 
Here $\chi^{e}_{j,-}(k_{F_j})$ and $\chi^{h}_{j,-}(k_{F_j})$ are given by
\begin{align}
&\chi^{e}_{1,-}(k_{F_1}) = \chi^{h}_{1,-}(k_{F_1}) = - \mathcal{T} \chi^{e}_{1,+}(k_{F_1}), \nonumber\\
&\chi^{e}_{2,-}(k_{F_2}) = \chi^{h}_{2,-}(k_{F_2}) = \mathcal{T} \chi^{e}_{2,+}(k_{F_1}),
\end{align}
where $\mathcal{T} = -i \tilde{\sigma}_y \Sigma_0 \mathcal{C}$ is the time reversal operator.
The coefficients $c_{j}(x)$ and $d_{j}(x)$ are obtained from   
\begin{align}
\begin{pmatrix}
\hbar v_j \left(-i \partial_x - k_{F_j}\right) & \Delta(x) e^{i \phi(x)} \\
\Delta(x) e^{-i \phi(x)} & \hbar v_j \left(i \partial_x + k_{F_j}\right)
\end{pmatrix} 
\begin{pmatrix} 
c_{j}\\
d_{j}
\end{pmatrix} = 
\varepsilon 
\begin{pmatrix} 
c_{j}\\
d_{j}
\end{pmatrix}, \nonumber
\end{align}
and the Andreev eigenstates are given by 
\begin{widetext}
\begin{align}
\Psi_{j-}(x) = \begin{cases}
\sqrt{\frac{\kappa_j}{N_j}} e^{(i k_{F_j}+\kappa_j) x} 
\begin{pmatrix}
e^{i \phi_L/2}~ \beta_j~ \chi^{e}_{j,-}(k_{F_j})\\
e^{-i \phi_L/2}~  \chi^{h}_{j,-}(k_{F_j})
\end{pmatrix} & \text{for  $x<0$}, \\
\sqrt{\frac{\kappa_j}{N_j}} 
\begin{pmatrix}
e^{i \phi_L/2 + i k^{e}_{xj} x}~ \beta_j~  \chi^{e}_{j,-}(k_{F_j})\\
e^{-i \phi_L/2 + i k^{h}_{xj} x}~ \chi^{h}_{j,-}(k_{F_j})
\end{pmatrix} & \text{for $0 \leq x \leq L$}, \\
\sqrt{\frac{\kappa_j}{N_j}} e^{(i k_{F_j}-\kappa_j) (x-L)} 
e^{-i(\phi_R - \phi_L)/2+i k^{e}_{xj} L}~\beta_j~
\begin{pmatrix}
e^{i \phi_R/2}~ \chi^{e}_{j,-}(k_{F_j})\\
e^{-i \phi_R/2}~ \beta_j~ \chi^{h}_{j,-}(k_{F_j})
\end{pmatrix} & \text{for $L < x$}.
\end{cases}
\end{align} 
\end{widetext}

\section{Andreev levels and current matrix elements of Josephson junctions in a 2DEG heterostructure} \label{app:2Dnanowire}

\begin{figure}[!t]
\centering
\includegraphics[width=\columnwidth]{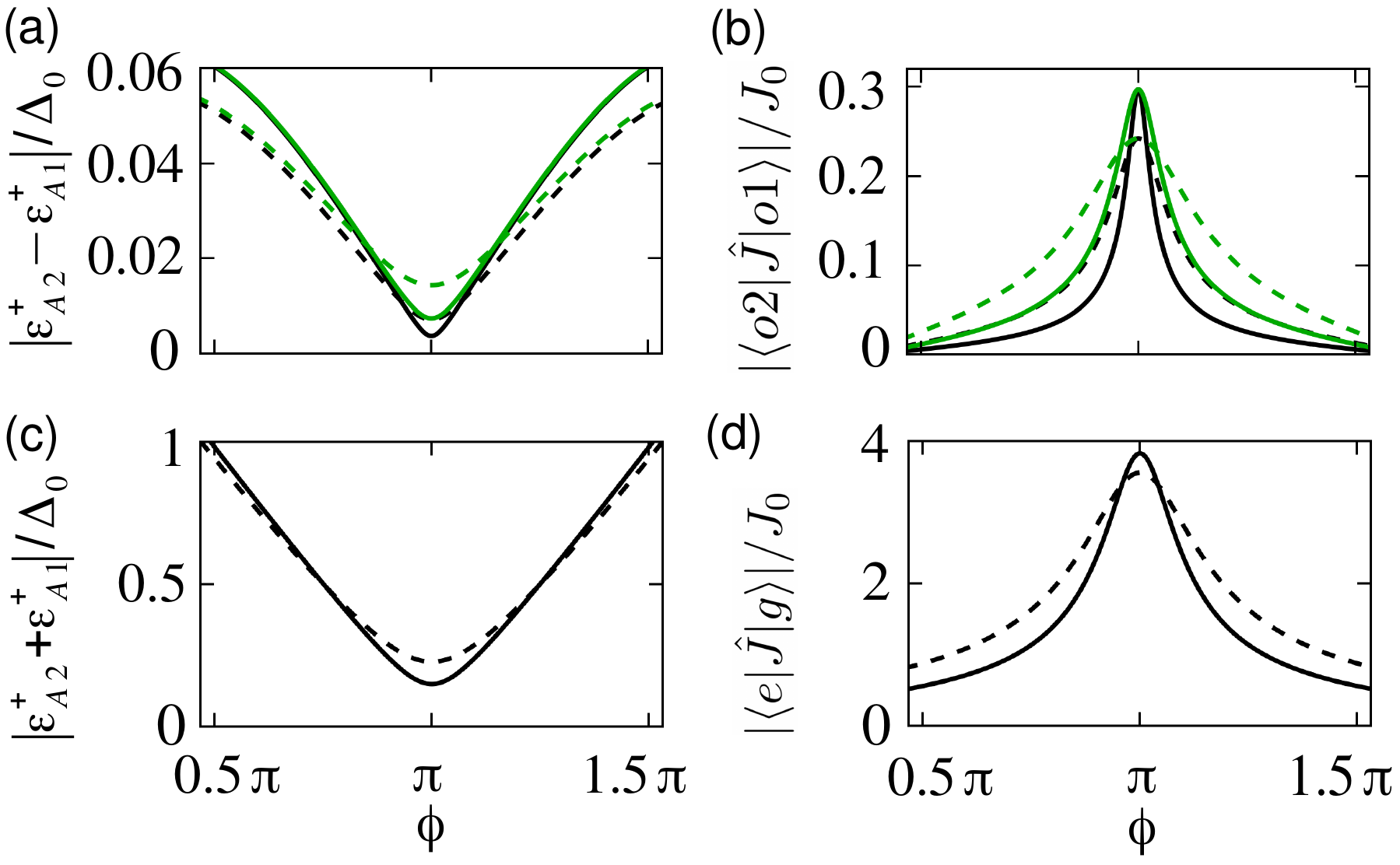}
\caption{Excitation spectra and matrix elements of the current operator in 2DEG-based Josephson junctions as a function of $\phi$ at $B_y =0$ for odd (a, b) and even (c, d) transitions.
The plots are drawn for different values of $\mu$ and $B_x$; 
$\mu = 1.4$ meV and $B_x = 50$ mT (black solid lines), $1$ meV and $50$ mT (black dashed), $1.4$ meV and $100$ mT (green solid), 
and $1$ meV and $100$ mT (green dashed). 
The other system parameters $\hbar \alpha = 40$ meV nm, $W = 200$ nm, $L = 300$ nm, $\Delta_0 = 165~\mu$eV, $g$-factor $= 12$, $U_0 = 16.5$ meV nm and $m=0.023~m_e$ are used. 
These values are the same as used in Fig.~\ref{Fig2:Result1}, 
except for a larger strength of the spin-orbit coupling. 
}
\label{App:Fig1}
\end{figure}    

\begin{figure}[!t]
\centering
\includegraphics[width=\columnwidth]{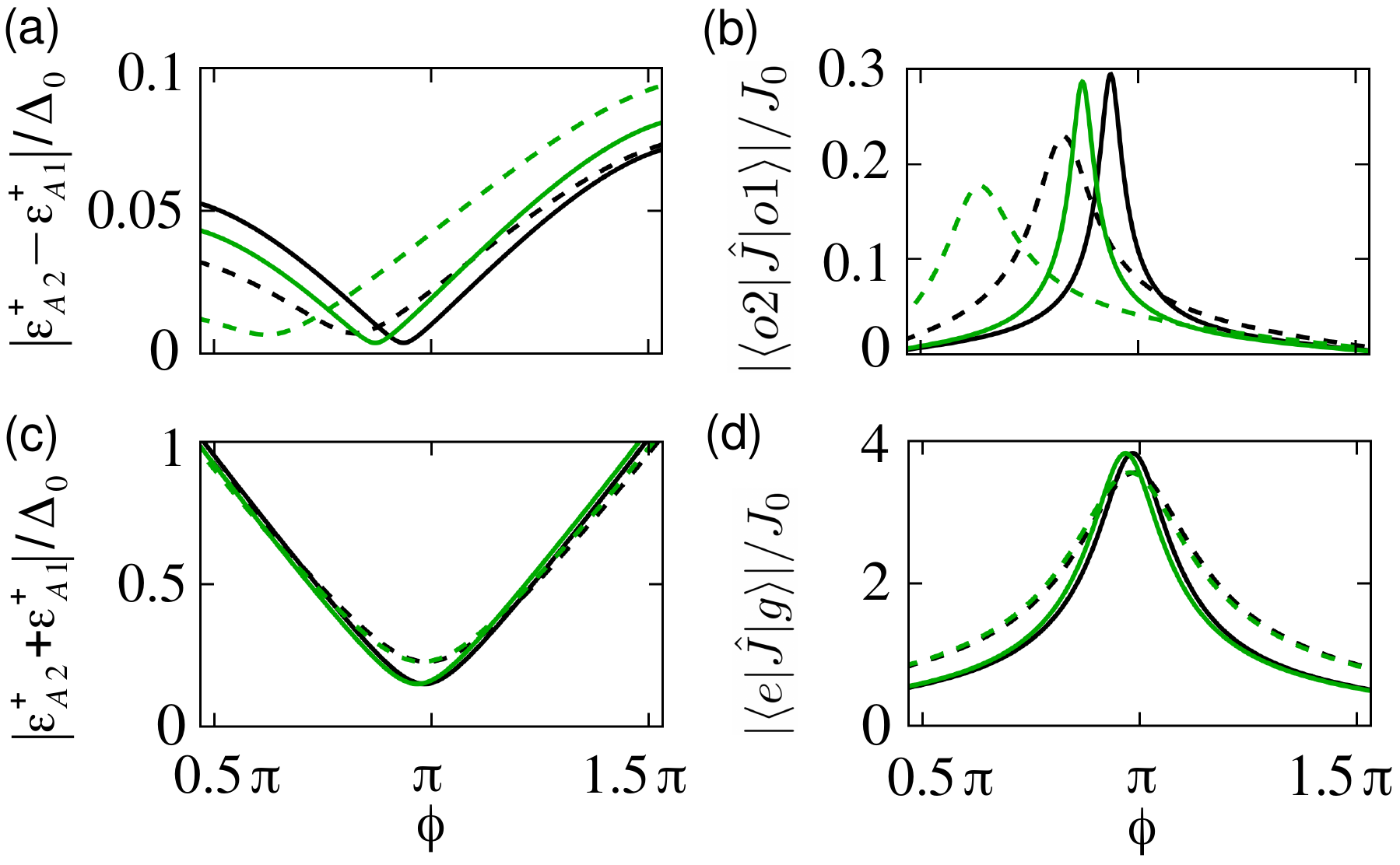}
\caption{Same plots as in Fig.~\ref{App:Fig1}, but for different values of $B_y$; $\mu = 1.4$ meV and $B_y = 10$ mT (black solid lines), $1$ meV and $10$ mT (black dashed), $1.4$ meV and $20$ mT (green solid), 
and $1$ meV and $20$ mT (green dashed). Here $B_x = 50$ mT is used.
}
\label{App:Fig2}
\end{figure}

In this appendix, we obtain an effective one-dimensional BdG Hamiltonian $H^{1D}_{\text{BdG}}$ for Josephson junctions in a 2DEG heterostructure where the electrons are confined in the $y$-direction with width $W_{2d}$ and free to move in the $x$-direction. The full Hamiltonian in this case is 
\begin{align}
H^{2d}_{\text{BdG}} = \left(H^{2d}_0-\mu\right) \tau_z + H_R \tau_z + H_Z + H_S,
\end{align}  
where $H^{2d}_0$, instead of the $H_0$ in Eq.~\eqref{H0} in the main text, is 
\begin{align}
H^{2d}_0 = \frac{p^2_x+p^2_y}{2m} + U_{b}(x)+U_{c}(y), 
\end{align}
where the hard-wall confinement potential $U_c(y)$ is defined as $U_c(y)=0$ for $0 < y < W_{2d}$ and $\infty$ otherwise. Here, $H_R$, $H_Z$, and $H_S$ are the same as given in Eq.~\eqref{H_BdG}.
We start by calculating transverse eigenvalues and their eigenstates by solving $p^2_y/(2m)+U_c(y)$. 
The eigenvalues are given by $E^{\perp}_n = (\hbar^2 \pi^2 n^2)/(2 m W^2_{2d})$ and corresponding eigenstates are 
\begin{align}
\phi^{\perp}_{ns}(y)=\frac{2}{\sqrt{W_{2d}}}~\text{sin}(n\pi y/W_{2d}) \chi_s,
\end{align}
where $n=1,2,...$ denote the indices for transverse subbands and 
$\chi_{\uparrow (\downarrow)} =  (1/\sqrt{2})(1, i (-i))^{T}$ are eigenstates of $\sigma_y$. 
Note that, different to the case of cylindrical nanowire with an harmonic confinement potential discussed in the main text, there is no degeneracy for the higher transverse subbands besides spin degeneracy. 
By projecting $H^{2d}_{\text{BdG}}$ onto the subspace spanned by the lowest two transverse subbands with $ns \in \{1\uparrow,1\downarrow,2\uparrow, 2\downarrow\}$ and by integrating out the $y$-coordinate, we have 
\begin{align}
H^{1D}_{\text{BdG}} = \left(H'^{2d}_0-\mu\right) \tau_z + H'^{2d}_R \tau_z + H'_Z + H_S, 
\end{align}
where $H'^{2d}_0$ and $H'^{2d}_R$ are given by
\begin{align}
H'^{2d}_0 &= \frac{p^2_x}{2m} + E^{\perp}_{2d+} + E^{\perp}_{2d-}\Sigma_z + U_{b}(x), \\
H'^{2d}_R &= -\alpha p_x \tilde{\sigma}_z - \eta_{2d} \tilde{\sigma}_y \Sigma_y, 
\end{align}
where $E^{\perp}_{2d\pm} = (E^{\perp}_1 \pm E^{\perp}_2)/2$.
The coefficient $\eta_{2d}$ in Eq.~\eqref{H'R} describes the coupling between the different transverse subbands with opposite spins, and is given by 
\begin{align}
\eta_{2d} &= \int^{W_{2d}}_{0} dy~ \phi^{\perp \dagger}_{1\uparrow}(y) \left(-i \hbar \alpha \frac{\partial}{\partial y} \sigma_x \right) \phi^{\perp}_{2\downarrow}(y) \nonumber\\
&= \frac{8\alpha \hbar}{3W_{2d}}. 
\end{align} 
The dispersion relation of the lowest subbands, which is obtained 
by solving $H'^{2d}_0 + H'^{2d}_R$ with $U_{b} = 0$, is computed as 
\begin{align}
E(k_x) = \frac{\hbar^2 k^2_x}{2m} + E^{\perp}_{2d+} 
- \sqrt{\left( E^{\perp}_{2d-} \mp \alpha \hbar k_x\right)^2 + \eta^2_{2d}}, 
\end{align} 
which is the same as in Eq.~\eqref{Dispersion}, except for replacing $E^{\perp}_{\pm}$ and $\eta$ by $E^{\perp}_{2d \pm}$ and $\eta_{2d}$, respectively. Extracting the parameters $v_{j=1,2}$ and $\theta_{j=1,2}$ from the dispersion and by using the mapped BdG Hamiltonian in Eq.~\eqref{PBdG}, we obtain the Andreev levels and current matrix elements for even and odd states. Fig.~\ref{App:Fig1} is plotted for the same parameter values as in Fig.~\ref{Fig2:Result1} except for a larger spin-orbit coupling, which shows the finite (no) dependence for the odd (even) transitions on $B_x$ as we have seen in Fig.~\ref{Fig2:Result1}, although the specific values of $|\varepsilon^{+}_{A2}\mp\varepsilon^{+}_{A2}|$, $\langle o2|\hat{J}| o1 \rangle$, and $\langle e |\hat{J}| g \rangle$ are different for the same system parameters due to the different dispersion relations. 
Furthermore, in Fig.~\ref{App:Fig2}, the same dependence on $B_y$ as shown in Fig.~\ref{Fig3:Result2} is presented such 
that the odd transitions significantly change by changing $B_y$ (Fig.~\ref{App:Fig2}(a) and (b)), while the even transitions 
is very weakly dependent on $B_y$ (Fig.~\ref{App:Fig2}(c) and (d)). 
We also checked the even and odd transition matrix elements for smaller spin-orbit coupling strength, $\hbar \alpha \sim 20$ meV nm, 
in this 2D geometry, and found that the matrix elements between the odd states are significantly smaller (almost two orders of magnitude smaller) for the smaller value of the $\alpha$ parameter. 
 
This comparison indicates that our findings - selectively tunable even and odd transitions by changing the system parameters like the Zeeman field, chemical potential, and transmission probability - are still valid in a 2DEG-based nanowire, and thus are independent of the nanowire geometry.

\section{Orbital effect of magnetic field in cylindrical nanowire Josephson junctions} \label{app:Orbital_effect}

In the main text, we neglected the orbital effect of a magnetic field $B_x$ which is characterized by a normalized magnetic flux $\Phi$, 
\begin{align}
\Phi = \frac{\pi B_x (W/2)^2}{h/e}, \label{Flux}
\end{align}
where $W$ is the diameter of the nanowire. 
In this appendix, we investigate the influence of the flux $\Phi$ to ABSs, and show that the account of the flux gives the corrections of the second order in $\Phi$ to the ABSs, and thereby the results obtained in Sec.~\ref{Model} and \ref{Current}, 
which are valid up to the first order in $B_x$, does not affected. 

To this end, we solve the following single-particle Hamiltonian associated with the transverse direction of the nanowire including the vector potential corresponding to $B_x$, 
\begin{align}
H^{\perp} = \frac{(p_y + e A_y)^2 + (p_z + e A_z)^2}{2m} + \frac{m \omega^2_0 (y^2+z^2)}{2m}, 
\end{align}     
where $(A_y,A_z) = (B_x/2) (-z,y)$. It is known as the Fock-Darwin Hamiltonian~\cite{Fock1928,Darwin1930}, and its eigenvalues are given by 
\begin{align}
E^{\perp}_{n_{r} n_{\theta}} = \hbar \Omega (2 n_{r} +1 + |n_{\theta}|) - \frac{\hbar \omega_c}{2} n_{\theta},
\end{align}
where $\Omega =\sqrt{\omega^2_{0} + \omega^2_{c}/4}$ and $\omega_{c} = e|B_x|/m$. $n_{r}= 0, 1, 2, ...$ is the quantum number in the radial direction and $n_{\theta} = 0, \pm 1, \pm 2,...$ is the angular momentum quantum number. 
From the definition of $W = 2 \sqrt{\hbar/(m \omega_0)}$ and Eq.~\eqref{Flux}, we can rewrite $\Omega$ and $\omega_c$ in terms of $\Phi$ as 
\begin{align}
\Omega &= \omega_0 \sqrt{1+ \Phi^2}, \nonumber\\
\omega_c &= 2 \omega_0 \Phi.
\end{align}
The eigenstates of the lowest three energies $E^{\perp}_{00}, E^{\perp}_{01}$, and $E^{\perp}_{0-1}$ are given by 
\begin{align}
\phi^{\perp}_{00s}(r) &= \frac{1}{\sqrt{2\pi}l_{\Omega}}e^{-r^2/(4 l^2_{\Omega})} \chi_s, \nonumber\\
\phi^{\perp}_{01s}(r,\theta) &= \frac{r e^{-i \theta}}{\sqrt{2\pi} \sqrt{2}l^{2}_{\Omega}}e^{-r^2/(4 l^2_{\Omega})} \chi_s, \nonumber\\
\phi^{\perp}_{0-1s}(r,\theta) &= \frac{r e^{i \theta}}{\sqrt{2\pi} \sqrt{2}l^{2}_{\Omega}}e^{-r^2/(4 l^2_{\Omega})} \chi_s,
\end{align} 
where $r=\sqrt{y^2+z^2}$, $\theta=\text{arctan}(z/y)$, and $l_{\Omega}=\sqrt{\hbar/(2m \Omega)}$. Similar to the procedure in Sec.~\ref{Model}, we project a Hamiltonian $p^2_x/(2m) - \alpha p_x \sigma_y + H^{\perp}+H_R$ onto the subspace spanned by the above eigenstates $\{\phi^{\perp}_{00\uparrow}, \phi^{\perp}_{00\downarrow},\phi^{\perp}_{01\uparrow}, \phi^{\perp}_{01\downarrow},\phi^{\perp}_{0-1\uparrow}, \phi^{\perp}_{0-1\downarrow} \}$, yielding a one-dimensional three-subband Hamiltonian $H^{1D}$,
\begin{gather}
H^{1D}~\psi = E~ \psi, \\
H^{1D} 
= 
\begin{pmatrix}
h_0 & -i \eta_b/\sqrt{2}~ \tilde{\sigma}_y & -i \eta_b/\sqrt{2}~ \tilde{\sigma}_y \\
i \eta_b/\sqrt{2}~ \tilde{\sigma}_y & h_{1} & 0 \\
i \eta_b/\sqrt{2}~ \tilde{\sigma}_y & 0 & h_{-1}
\end{pmatrix},  
\end{gather}
where $\psi = (\psi_{0\uparrow},\psi_{0\downarrow},\psi_{1\uparrow},\psi_{1\downarrow},\psi_{-1\uparrow},\psi_{-1\downarrow})$ and 
\begin{align}
\eta_b = \frac{\alpha \hbar}{2 l_{\Omega}} = \alpha \sqrt{\frac{m\hbar \omega_0}{2}}(1+\Phi^2)^{1/4} = \eta (1+\Phi^2)^{1/4}.
\end{align}  
where $\eta$ is given in Eq.~\eqref{Eta} in the main text. The diagonal elements of $H^{1D}$ are given by 
\begin{align}
h_{j} = \frac{\hbar^2 k^2_x}{2m} - \alpha \hbar k_x \tilde{\sigma}_z + E^{\perp}_{0j},
\end{align}
where $j\in \{0,1,-1\}$. We expand the parameters $\Omega$ and $\eta_b$ in $\Phi$, and retain up to the second order in $\Phi$. Then the dispersion relation for the lowest subbands is 
\begin{align}
E(k_x, \Phi) = E^{(0)}(k_x) + E^{(1)}(k_x) \Phi^2 + \mathcal{O}(\Phi^4), 
\end{align}
where $E^{(0)}(k_x)$ and $E^{(1)}(k_x)$ are given by
\begin{align}
E^{(0)}(k_x) = \frac{\hbar^2 k^2_x}{2m} + \frac{3\hbar \omega_0}{2} 
-\sqrt{\left(\frac{\hbar \omega_0}{2} + s' \alpha \hbar k_x \right)^2 + \eta^2},
\end{align}
and 
\begin{widetext}
\begin{align}
E^{(1)}(k_x) = \frac{\hbar \omega_0}{2} + \frac{s' \alpha \hbar k_x (\hbar^2 \omega^2_0-\eta^2/2) 
- (\hbar^2 \omega^2_0 + \eta^2/2) \left[\hbar^2 k^2_x/(2m)+\hbar \omega_0 - E^{(0)}(k_x)\right]}{\left[\hbar^2 k^2_x/(2m)+s' \alpha \hbar k_x + 2 \hbar \omega_0 - E^{(0)}(k_x)\right]^2+\eta^2},
\end{align}
\end{widetext}
where $s'=\pm 1$ are eigenvalues of $\tilde{\sigma}_z$ and hence distinguish the different spin subbands.
Note that $E^{(0)}(k_x)$ is the same as in Eq.~\eqref{Dispersion}, and that the leading order correction to the dispersion relation is the second order in $\Phi$.

For further comparison with $H'_0 + H'_R$ in Eqs.~\eqref{H'0} and \eqref{H'R} in the main text, we perform a transformation to $H^{1D}$ as 
\begin{align}
\begin{pmatrix}
\psi_{0s}\\
\psi_{1s}\\
\psi_{-1s}
\end{pmatrix}
\rightarrow
\begin{pmatrix}
\psi_{0s}\\
\psi'_{1s}\\
\psi'_{-1s}
\end{pmatrix},
\end{align}
where $\psi'_{1s} = (\psi_{1s}+\psi_{-1s})/\sqrt{2}$ and $\psi'_{-1s} = (\psi_{1s}-\psi_{-1s})/\sqrt{2}$, followed by eliminating the $\psi'_{-1s}$ components, yielding 
\begin{align}
H'^{1D} 
= 
\begin{pmatrix}
h_0 & -i \eta_b \tilde{\sigma}_y  \\
i \eta_b \tilde{\sigma}_y & h'_{1} + \delta h'_1 \Phi^2
\end{pmatrix},
\end{align}
where $h'_1$ and the self energy correction $\delta h'_1$ from the $\psi'_{-1s}$ components are given by 
\begin{align} 
h'_1 &= \frac{\hbar^2 k^2_x}{2m} - \alpha \hbar k_x \tilde{\sigma}_z + 2 \hbar \Omega, \nonumber\\
\delta h'_1 &= -\frac{\hbar^2 \omega^2_0}{h'_1 - E}
\end{align}
It is easy to check that $H'^{1D} = H'_0 + H'_R$ if $\Phi = 0$ and $U_b =0$. 
By comparing the dispersion relations of $H^{1D}$ and $H'^{1D}$ order by order in $\Phi$, we find the form of $\delta h'_1$, which shows consistency up to $\Phi^2$-order, as 
\begin{align}
&\delta h'_1 = E^{(1)}(k_x) - \hbar \omega_0 \nonumber\\
&\hspace{10pt}+ \frac{\left(E^{(1)}(k_x) - \frac{\hbar \omega_0}{2}\right) \left(\frac{\hbar^2 k^2_x}{2m}+\alpha \hbar k_x \tilde{\sigma}_z+2 \hbar \omega_0 \right) + \frac{\eta^2}{2}}{\frac{\hbar^2 k^2_x}{2m}-\alpha \hbar k_x \tilde{\sigma}_z+ \hbar \omega_0 - E^{(0)}(k_x)}. 
\end{align}
As a result, the $\Phi$ induced corrections to the $H'_0 + H'_R$ are found as
\begin{align}
H'^{1D} = H'_0 + H'_R + 
\begin{pmatrix}
\frac{\hbar \omega_0}{2} & -i \frac{\eta}{4}\tilde{\sigma}_y \\
i \frac{\eta}{4}\tilde{\sigma}_y & \hbar \omega_0 + \delta h'_1
\end{pmatrix} 
\Phi^2 + \mathcal{O}(\Phi^4).  \label{H_with_flux}
\end{align}
This correction term would lead to the $\Phi^2$-order corrections to the Fermi velocities and the wave functions of electrons in the lowest subbands. Therefore, our perturbative results for ABSs and current operator matrix elements given in Sec.~\ref{Current} up to the first order in $B_x$ are still valid, provided that $\Phi \ll 1$. For larger $\Phi$ values, Eq.~\eqref{H_with_flux} would allow to calculate the effect on all the results in the main text with an accuracy of $\mathcal{O}(\Phi^4)$.

\end{document}